%% file: main.tex
\documentclass[10pt,twocolumn,letterpaper]{article}

\usepackage{cvpr}              
\usepackage{algorithm}       
\usepackage{algorithmic}     
\usepackage{amsmath}         
\usepackage{amsmath} 
\usepackage{multirow}
\usepackage{soul}     
\usepackage{comment}

\usepackage{amsmath}         
\usepackage{amssymb}        
\usepackage{amsthm}         
\usepackage{mathtools}      

\usepackage{bm}             
\usepackage{enumitem}       
\usepackage{mathrsfs}       
\usepackage{bbm}            

\usepackage{tikz}
\usetikzlibrary{arrows.meta}
\usetikzlibrary{decorations.pathreplacing}
\usepackage{array}          
\usepackage{upgreek}

\newcommand{\frameworkname}{UIBDiffusion}


\definecolor{cvprblue}{rgb}{0.21,0.49,0.74}
\usepackage[pagebackref,breaklinks,colorlinks,allcolors=cvprblue]{hyperref}

\def\paperID{11881} 
\def\confName{CVPR}
\def\confYear{2025}

\title{UIBDiffusion: Universal Imperceptible Backdoor Attack for Diffusion Models}

\author{Yuning Han\textsuperscript{1}\thanks{Equal contribution.}~~~Bingyin Zhao\textsuperscript{2}\footnotemark[1]~~~Rui Chu\textsuperscript{3}~~~Feng Luo\textsuperscript{4}~~~
Biplab Sikdar\textsuperscript{2}~~~
Yingjie Lao\textsuperscript{3}\thanks{Corresponding author.}\\\\
\textsuperscript{1}Columbia University \quad \textsuperscript{2}National University of Singapore \quad \textsuperscript{3}Tufts University \quad \textsuperscript{4}Clemson University
}

\begin{document}
\maketitle
\input{sec/0_abstract}    
\input{sec/1_intro}

\input{sec/2_background}
\input{sec/3_method}
\input{sec/4_Experiments}
\input{sec/5_conclusion}

\clearpage
{
    \small
    \bibliographystyle{unsrt}
    \bibliography{main}
}

\clearpage
\input{sec/6_Appendix}


\end{document}

%% file: sec/0_abstract.tex
\begin{abstract}
Recent studies show that diffusion models (DMs) are vulnerable to backdoor attacks. Existing backdoor attacks impose unconcealed triggers (e.g., a gray box and eyeglasses) that contain evident patterns, rendering remarkable attack effects yet easy detection upon human inspection and defensive algorithms. 
While it is possible to improve stealthiness by reducing the strength of the backdoor, doing so can significantly compromise its generality and effectiveness. In this paper, we propose \frameworkname, the \ul{u}niversal \ul{i}mperceptible \ul{b}ackdoor attack for diffusion models, which allows us to achieve superior attack and generation performance while evading state-of-the-art defenses. We propose a novel trigger generation approach based on universal adversarial perturbations (UAPs) and reveal that such perturbations, which are initially devised for fooling pre-trained discriminative models, can be adapted as potent imperceptible backdoor triggers for DMs. We evaluate \frameworkname \space on multiple types of DMs with different kinds of samplers across various datasets and targets. Experimental results demonstrate that \frameworkname \space brings three advantages: 1) \textit{Universality}, the imperceptible trigger is universal (i.e., image and model agnostic) where a single trigger is effective to any images and all diffusion models with different samplers; 2) \textit{Utility}, it achieves comparable generation quality (e.g., FID) and even better attack success rate (i.e., ASR) at low poison rates compared to the prior works; and 3) \textit{Undetectability}, \frameworkname \space is plausible to human perception and can bypass Elijah and TERD, the SOTA defenses against backdoors for DMs. We will release our backdoor triggers and code.

\end{abstract}

%% file: sec/1_intro.tex
\section{Introduction}
\label{sec:intro}


Diffusion models (DMs)~\cite{ho2020denoising,DBLP:conf/iclr/SongME21,DBLP:conf/cvpr/RombachBLEO22} have emerged as the state-of-the-art generative paradigm in recent years and achieved unprecedented success in image~\cite{DBLP:conf/iccv/PeeblesX23,DBLP:conf/icml/EsserKBEMSLLSBP24}, video~\cite{ho2022video, lu2022dpm, jiang2024videobooth} text~\cite{li2022diffusion,xu2023versatile,zhang2023adding} and audio~\cite{jeong2021diff, huang2022fastdiff, kim2022guided, popov2021grad, wang2023audit, shen2023difftalk} synthesis. The success is highly attributed to training the foundation model on large-scale datasets~\cite{DBLP:conf/icml/EsserKBEMSLLSBP24}.  Such a practice places DMs at risk of being compromised by data poisoning attacks~\cite{li2016data,zhao2018data, alfeld2016data}, where adversaries contaminate the clean training dataset with a small fraction of malicious samples and undermine model performance. This imposes a critical security concern on the practical deployment of DMs.

\begin{figure*}[htbp]
    \centering
    \includegraphics[width=0.9\linewidth]{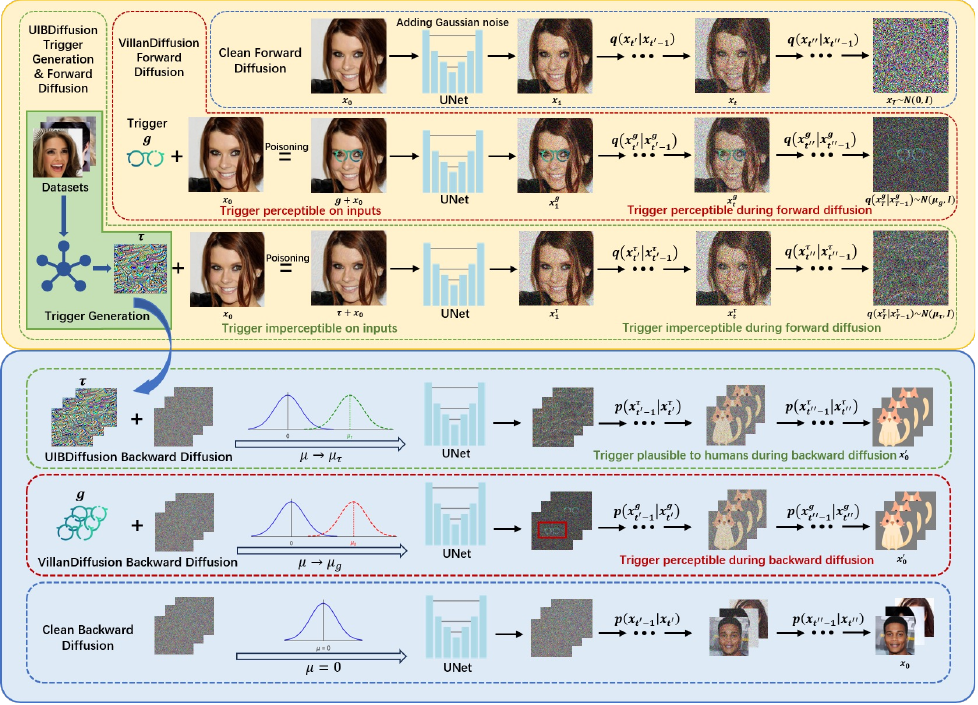}
    \caption{Illustrations of the forward diffusion process (top block with the yellow background) and backward diffusion process (bottom block with the blue background) of a clean diffusion model (blue dash line), VillanDiffusion~\cite{chou2024villandiffusion} (red dash line) and \frameworkname \space(Ours, green dash line). The \frameworkname \space trigger ($\uptau$) is plausible to humans in all phases from data poisoning to forward diffusion and backward diffusion while the glasses trigger ($\mathbf{g}$) in prior works is perceptible. \frameworkname \space trigger is highly effective since it introduces a similar distribution shift as the glasses trigger, which secures the attack performance during the backward diffusion process. However, it is hard to detect as the trigger does not possess specific patterns and cannot be inverted by existing defensive algorithms. We empirically verify the effectiveness in Section~\ref{sec:exp}.} 
    \label{fig:flow}
    \vspace{-0.8em}
\end{figure*}

Backdoor attacks are emerging threats to DMs. Recent studies~\cite{chou2023backdoor,chen2023trojdiff,chou2024villandiffusion} show that DMs are vulnerable to such attacks, which implant backdoor into victim models by training them on poisoned datasets and manipulate their behavior through inputs stamped with predefined triggers. Existing backdoor attacks for DMs usually employ unconcealed triggers such as a gray box, an image of Hello Kitty, and eyeglasses in the forward and reversed (i.e., sampling) diffusion process, as shown in Fig.~\ref{fig:flow}. While such trigger designs achieve superior attack performance and preserve the decent generation ability of victim models, they are easy to detect via human inspection and advanced defensive algorithms as they have obvious patterns that can be precisely remodeled via reverse engineering~\cite{an2024elijah,DBLP:conf/icml/MoHLL024}. Motivated by this observation, we aim to design a general, powerful yet stealthy trigger. However, even for the traditional backdoor attacks against discriminative models (e.g., image classifiers)~\cite{gu2017badnets,DBLP:journals/corr/abs-1712-05526,Trojannn}, it is challenging to acquire such a trigger since it either requires complex and carefully designed generators to generate image-specific or model-specific triggers~\cite{DBLP:conf/ccs/YaoLZZ19,DBLP:conf/eccv/LiuM0020,li2021invisible,DBLP:conf/iclr/NguyenT21,DBLP:conf/aaai/0005LMZ21} or suffers from a relatively low attack success rate (ASR)~\cite{DBLP:conf/aaai/SahaSP20,DBLP:conf/nips/SouriFCGG22}, or fails to evade modern defenses~\cite{DBLP:conf/nips/LiLKLLM21,DBLP:conf/aisec-ws/ZhaoW24,DBLP:conf/sp/WangYSLVZZ19,DBLP:conf/nips/WuW21,DBLP:conf/iclr/LiLKLLM21}. This raises an interesting question in backdoor attacks against DMs: \textit{Is there a trigger simultaneously possessing generality, effectiveness, and stealthiness for backdoor attacks against DMs?}


Driving by the question, in this work, we propose \frameworkname, the universal imperceptible backdoor triggers for DMs that bring the following advantages: 1) \textit{Universality}: the trigger is image-agnostic and model-agnostic so that it can be applied to arbitrary images and DMs; 2) \textit{Utility}: the trigger can achieve high ASR while maintaining the generation capability of DMs; 3) \textit{Undetectability}: the trigger is plausible to humans and capable of circumventing state-of-the-art defenses. Inspired by the adversarial perturbation~\cite{DBLP:journals/corr/SzegedyZSBEGF13} originally designed for attacking pre-trained image classifiers, we reveal that universal adversarial perturbations~\cite{moosavi2017universal} (UAPs), a variant of adversarial perturbations, naturally hold the aforementioned properties and can be adapted as the \frameworkname \space triggers. First, UAPs are image-agnostic and model-agnostic. Second, we demonstrate that they are effective triggers because they introduce a similar distribution shift as prior works, yet they are stealthy because they do not possess clear patterns and are hard to estimate by existing trigger inversion algorithms. Moreover, they are trivial and imperceptible noises that humans can barely identify, as shown in Fig.~\ref{fig:flow}. 
Our contributions are summarized as follows:
\begin{itemize}
    \item To the best of our knowledge, \frameworkname \space is the first imperceptible backdoor attack against diffusion models.

    \item We discover that adversarial perturbations designed for \textit{discriminative models} can be adapted as backdoor triggers for \textit{generative models}. We reveal why such triggers can achieve high ASR and evade SOTA defenses.

    \item We propose a practical trigger generation approach adapted from UAP, which is specifically designed to enhance universality, attack effectiveness and accessibility of triggers.

    \item We experimentally demonstrate that \frameworkname \space is highly effective towards multiple diffusion models with different samplers and achieves comparable generation quality on benign input compared to clean DMs and superior attack success rate and input stealthiness compared to prior works. Moreover, 
    
    \item While existing works face a trade-off between utility and undetectability, we empirically show that \frameworkname \space triggers are highly resistant to the SOTA trigger inversion-based defenses, effectively bypassing Elijah~\cite{an2024elijah} and TERD~\cite{DBLP:conf/icml/MoHLL024}, the state-of-the-art defenses against backdoors on DMs.

\end{itemize}


%% file: sec/2_background.tex
\section{Related Works}
\label{sec:background}

\subsection{Diffusion Models} 
Diffusion models are latent variable models composed of two processes. The forward diffusion process imposes Gaussian noise on input data in each time step along the Markov chain and progressively diffuses the data distribution to an isotropic Gaussian distribution. The backward diffusion process is a reversed Markov chain that gradually denoises the sampled noise from the Gaussian distribution to the input data distribution. Mainstream diffusion models include DDPM~\cite{ho2020denoising}, LDM~\cite{DBLP:conf/cvpr/RombachBLEO22} and NCSN~\cite{DBLP:conf/nips/SongE19,DBLP:conf/iclr/0011SKKEP21,DBLP:conf/nips/0011E20}, etc. Despite the excellent performance, DMs suffer from slow sampling during the backward process. There exists a line of works that leverage different techniques such as generalized non-Markovian chain and Ordinary Differential Equation (ODE) to improve the samplers and accelerate the generation process: DDIM~\cite{DBLP:conf/iclr/SongME21}, DEIS~\cite{DBLP:conf/iclr/ZhangC23}, UniPC~\cite{DBLP:conf/nips/ZhaoBR0L23}, Heun~\cite{DBLP:conf/nips/KarrasAAL22}, DPM-Solver~\cite{DBLP:conf/nips/0011ZB0L022} and DPM-Solver++~\cite{lu2022dpm}. We comprehensively evaluate \frameworkname \space on these models and samplers to showcase the utility of our trigger. 

\subsection{Backdoor Attacks and Defenses on Diffusion Models}
\textbf{Attacks.} Backdoor attacks~\cite{gu2017badnets,DBLP:journals/corr/abs-1712-05526,Trojannn}, originally explored in the context of discriminative models such as image classifiers, induce models to make wrong predictions on triggered images. Although having been extensively studied for discriminative models, backdoor attacks are under-explored in diffusion models. BadDiffusion~\cite{chou2023backdoor}, TrojDiff~\cite{chen2023trojdiff} and VillanDiffusion~\cite{chou2024villandiffusion} are pioneering works in this field. Unlike traditional backdoor attacks, those targeting DMs aim to force models to synthesize a targeted image upon malicious inputs (i.e., noises with triggers) while generating normal images upon benign inputs (i.e., isotropic Gaussian noises). BadDiffusion shows how to backdoor DMs with unconcealed triggers such as a gray box and a stop sign. TrojDiff proposes to use a whole image (e.g., Hello Kitty) as the trigger and demonstrates that backdoor can be generalized to DDIM and different adversarial goals. However, both attacks are limited to certain DMs (e.g., DDPM) and samplers (e.g., ancestral and DDIM). VillanDiffusion resolves the limitations and offers a unified framework for all kinds of DMs and samplers. Other works show backdoor can be injected to text-to-image DMs and activated by text prompts~\cite{DBLP:conf/mm/ZhaiDSPF023,DBLP:journals/corr/abs-2305-10701,DBLP:conf/iccv/StruppekHK23}. However, these attacks draw less attention from defenders as they can eliminate the backdoor by cleansing the text encoder. Thus, our attack focuses on the primary defensive objective of the defender: pixel level (i.e., image) trigger generation.

\noindent \textbf{Defenses.} Backdoor attacks against DMs are shown to be immune to defenses designed for traditional image classifier backdoors~\cite{DBLP:conf/aaai/ChenCBLELMS19,DBLP:conf/iccv/ZengPMJ21,DBLP:conf/nips/WuW21, chou2023backdoor,chou2024villandiffusion}. The reason is that DMs take sampled noises as inputs while image classifiers take natural images as inputs~\cite{DBLP:conf/icml/MoHLL024}. Elijah~\cite{an2024elijah} is the first defensive algorithm specifically proposed against DM backdoors. It presents a trigger inversion scheme that can approximate the triggers and detect and remove the injected backdoor accordingly. TERD~\cite{DBLP:conf/icml/MoHLL024} employs a similar trigger estimation strategy and improves the trigger inversion process via a unified loss and trigger refinement. Based on the reconstructed trigger, it subsequently proposes to measure the KL divergence between benign and reversed distribution to detect the backdoor injected in models. Elijah and TERD achieve remarkable detection performance on existing DM backdoors. Thus, we evaluate \frameworkname \space against them and show that \frameworkname \space can bypass both defenses.

\subsection{Adversarial Perturbation}
The concept of adversarial perturbation is proposed in~\cite{DBLP:journals/corr/SzegedyZSBEGF13}, demonstrating that deep neural networks are vulnerable to small, additive noise. These perturbations are typically subtle and imperceptible, hence are widely used to craft adversarial examples~\cite{DBLP:journals/corr/GoodfellowSS14,DBLP:conf/iclr/KurakinGB17a,DBLP:conf/ijcai/XiaoLZHLS18} to fool image classifiers. Later,  \cite{moosavi2017universal} devised universal adversarial perturbation (UAP), a model-agnostic and image-agnostic variant of adversarial perturbation. We find that UAP is a particularly competitive triggers for backdoor attacks against DMs due to their imperceptibility, generalization property and accessibility. 

%% file: sec/3_method.tex
\section{\frameworkname}
\label{sec:method}

\subsection{Threat Model}
We adopt the same threat model and attack scenario presented in the existing backdoor attacks on DMs~\cite{chou2023backdoor,chen2023trojdiff,chou2024villandiffusion}, where adversaries publish the backdoored models on websites such as HuggingFace, and users can acquire the pre-trained models via third-party sources.  Adversaries are expected to be able to inject the backdoor and verify the attack and generation performance before releasing the models. Users are supposed to get access to the backdoor models and the subset of clean training data and test the model utility on their side. Note that in practice, \frameworkname \space applies to a more stringent threat model where users can access the entire training dataset (i.e., contaminated dataset with poisoned samples) and apply any defenses before model deployment as the trigger is imperceptible and undetectable.

\subsection{Inject Backdoor in Diffusion Models}
Prior attacks leveraged similar backdoor injection mechanisms with slight differences in sampler modifications~\cite{chou2023backdoor,chen2023trojdiff,chou2024villandiffusion}. \frameworkname \space employs the scheme presented in VillanDiffuision~\cite{chou2024villandiffusion} since it provides a unified framework to implant backdoors in various types of diffusion models and samplers and achieves outstanding performance. We briefly introduce the injection process in this section.

\noindent \textbf{Clean diffusion process.} Clean diffusion consists of forward and backward processes. The forward diffusion process diffuses images $x$ from data distribution $q(\text{x}_0)$ to the Gaussian distribution $\mathcal{N}(0,\textbf{I})$ by progressively adding noise in each time step along a Markov chain. The process can be mathematically expressed as: $q(x_t | x_0) = \mathcal{N}(\alpha(t) x_0, \beta^2(t) \textbf{I})$ with (1) $ q(x_{T_{\text{max}}}) \approx \mathcal{N}(0,\textbf{I})$ , (2) $ q(x_{T_{min}}) \approx q(x_0)$ and $x_t, t \in [T_{\text{min}}, T_{\text{max}}]$, where $T_{min}, T_{max}$ are the first and last timestep, and $\alpha(t), \beta(t)$ are parameterized variables introduced in the original diffusion model work~\cite{ho2020denoising}.

Backward diffusion process reverse the Markov chain and generate images from a sampled Gaussian noise by gradually removing noises in each timestep and can be expressed as:  $q(x_t | x_{t-1}) = \mathcal{N}(k_t x_{t-1}, w_t^2 \textbf{I})$, and $x_t(x_0,\epsilon)=\hat{\alpha}(t)x_0+\hat{\beta}(t)\epsilon$, where $\epsilon\sim \mathcal{N}(0,\textbf{I})$ and $k_t$, $w_t$,  $\hat{\alpha}(t),\hat{\beta}(t)$ are the same parameterized variables for the reverse process.

\noindent\textbf{Backdoor and trigger injection.} We employ the VillanDiffusion framework for backdoor injection and modify the trigger injection process in data poisoning. VillanDiffusion proposes a unified backdoor injection framework by introducing the loss function shown in Equation~\ref{equ:villan_loss}, where $\mathcal{L}_c$ and $\mathcal{L}_p$ are used to optimize the original generation and backdoor attack tasks, respectively.  $r$ is the poisoned data, $y$ is the backdoor target, $\zeta$ is a parameter to control backdoor against different samplers (e.g., ODE and SDE), and $\eta_c,\eta_p$ are weights to balance $\mathcal{L}_c$ and $\mathcal{L}_p$.

\begin{equation}
\begin{aligned}
& \mathcal{L}_{\theta}(\eta_c, \eta_p, x, t, \epsilon, r, y, \xi) = \\
& \eta_c \mathcal{L}_c(x, t, \epsilon) + \eta_p \mathcal{L}_p(x, t, \epsilon, r, y, \zeta),
\end{aligned}
\label{equ:villan_loss}
\end{equation}

We then explain the difference of the trigger injection process between VillanDiffusion and \frameworkname. VillanDiffusion injects the backdoor trigger follow the method in Equation~\ref{equ:villan_backdoor_injection}, where $\textbf{M}$ is a mask and $g$ is the trigger with specific patterns (e.g., gray box and a pair of glasses), and $\odot$ is the element-wise multiplication.
\begin{equation}
\begin{aligned}
    r(x,g) =\textbf{M}\odot g+(1-\textbf{M})\odot x.
\end{aligned}
\label{equ:villan_backdoor_injection}
\end{equation}
To better algin with the noise adding scheme in the adversarial perturbation, we inject our trigger using the approach in Equation~\ref{equ:our_trigger_injection}, where $\uptau$ is the imperceptible trigger synthesized by our proposed trigger generator and $\varepsilon$ is a variable that controls the strength of the added trigger. In contrast to triggers with recognizable patterns, the \frameworkname \space triggers are generated by a trainable generator with subtle perturbations and hard-to-detect patterns, which makes them effective and stealthy. The backdoor and trigger injection process is summarized in Algorithm~\ref{alg:diffusion}.

\begin{equation}
 r(x,\uptau) = x + \varepsilon \odot \uptau
 \label{equ:our_trigger_injection}
\end{equation}




\input{Algorithm/Diffusion_Alg}
 
Our goal is to force the backdoored DMs to learn a shortcut projection from poisoned image distribution 
to the target image distribution. 
We can formulate the backdoored DM's forward process as $q(x'_t|x_0') = \mathcal{N}(\hat{\alpha}(t) x_0' + \hat{\rho}(t) r, \hat{\beta}^2(t) \textbf{I})$ in which: (1) $q(x'_{T_{\text{max}}}) \approx \mathcal{N}(r, \hat{\beta}^2(T_{\text{max}}) \textbf{I})$ and (2) $q(x'_{T_{\text{min}}} ) \approx \mathcal{N}(x_0', 0)$,  where $x_0$ is the backdoor target,  $\hat{\rho}(t)$ is a similar parametrized variable as $\hat{\alpha}(t)$ and $\hat{\beta}(t)$.
In the backdoor backward diffusion process, we can denote similar Markov diffusion process as $q(x_t'| x_{t-1}') = \mathcal{N}(k_t x_{t-1}' + h_t r, w_t^2 \textbf{I})$, 
where $h_t$ is a similar parametrized variables as $\hat{\alpha}(t)$ and $\hat{\beta}(t)$. Next, we describe the detailed proposed trigger generation approach of \frameworkname.

\subsection{\frameworkname \space Trigger Generation}
\label{sec:trigger_generation}
Recall that our goal is to design a general, effective, and stealthy trigger for backdoor attack against DMs. Due to the imperceptibility, generalization, and effectiveness properties of UAPs, they can be adapted as the \frameworkname \space trigger. We first briefly introduce the mechanism and generation process of UAPs. For a given image classifier $\mathcal{C}$ and a data distribution $q$, UAP algorithms aim to find a perturbation vector $v$ such that $\mathcal{C}(x+v)\neq\mathcal{C}(x)$ for most $x\sim q$ (i.e., image agnostic). The original UAP algorithm~\cite{moosavi2017universal} achieves this goal by measuring the minimal perturbation that pushes the perturbed input to the decision boundary, which can be mathematically expressed as:

\begin{equation}
    \Delta{v_i} = \mathop{\arg\min}\limits_{d}||d||_2 \quad \textbf{s.t.} \quad \mathcal{C}(x_i+v+d)\neq\mathcal{C}(x_i),
\end{equation}
where $i$ is the image index of a subset of the dataset selected for the computation. The final perturbation is updated as $\mathcal{P}_\xi(v+\Delta v_i)$, where $\mathcal{P}_\xi(\cdot)$ is a projection function and $\xi$ is the perturbation budget. Note that in the context of DMs backdoor, the perturbation $v$ is actually the imperceptible trigger $\uptau$. We use $\uptau$ for consistent notation from now on to avoid confusion.

However, the original UAP algorithm using DeepFool~\cite{moosavi2016deepfool} sometimes yields less effectiveness, and we experimentally observe it renders a relatively lower ASR as the backdoor trigger for DMs. Thus, we propose a novel trigger generation approach inspired by the GUAP~\cite{zhang2020generalizing}, an enhanced version of UAP that is more effective and robust. Our trigger generation process adopts the advantage of both additive (i.e., $l_{\infty}$-bounded) and non-additive (i.e., spatial transformation) perturbations, where the additive perturbation is adopted as the \frameworkname \space trigger $\uptau$ and coordinately optimized with the non-additive perturbation. We denote the non-additive noise as $f$. We employ a generator $\mathcal{G}_\gamma$ (parameterized by $\gamma$), which is akin to the generator in generative adversarial nets (GAN)~\cite{goodfellow2014generative} that take a latent noise $z$ as input, and generate $f$\ and trigger $\uptau$, as shown in lines 3, 4 and 5 in Algorithm~\ref{alg:backdoor_uib_uap}. In contrast to the GAN that optimizes the generator under the guide of a discriminator, we update our trigger generator under the guide of an image classier $\mathcal{C}$ (e.g., VGG and ResNet, etc.), which indicates whether the adversarial perturbation is effective or not, as expressed in Equation~\ref{equ:tau_loss}, where $\mathcal{H}(\cdot)$ is the cross entropy loss and $\otimes$ represents for the spatial transformation operation. Meanwhile, the non-additive perturbation is updated according to the loss function expressed in Equation~\ref{equ:f_loss}, where $\mathcal{S}(\cdot)$ is summed max pooling values of $x$ using a kernel of size 4 during the spatial transformation and $n$ is the number of images in the selected subset of the dataset. By incorporating such a design, we can progressively optimize the generator and hence improve the effectiveness of the generated \frameworkname \space trigger, as described in lines 5, 7 and 8 in Algorithm~\ref{alg:backdoor_uib_uap}, respectively. The completed trigger generation procedures are presented in Algorithm~\ref{alg:backdoor_uib_uap} and the concrete architecture of $\mathcal{G}_\gamma$ and flow illustration are provided in the Appendix as they are similar to GAN.

\begin{equation}
    \mathcal{L}_{\mathcal{G}} = -\mathcal{H}(\mathcal{C}(x\otimes f+\uptau), \mathcal{C}(x)), 
\label{equ:tau_loss}
\end{equation}

\begin{equation}
    \mathcal{L}_{f} = \max_{i}{\sqrt{\frac{1}{n}\sum^n_i (\mathcal{S}(x_i,f))}}
\label{equ:f_loss}
\end{equation}


 

\input{Algorithm/Algorithm}

\begin{figure}
    \centering
    
    \includegraphics[width=1\linewidth]{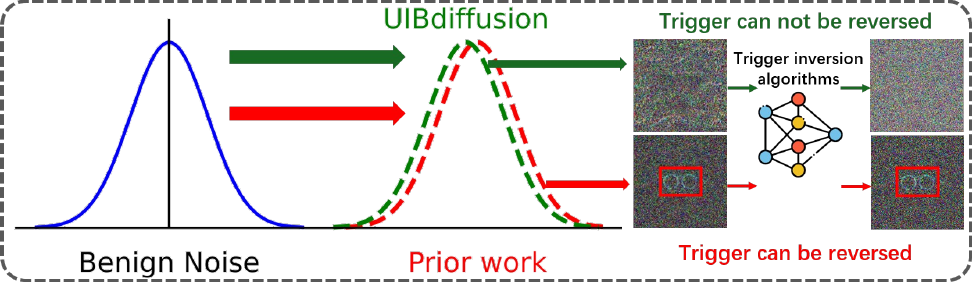}
    \caption{Illustration of trigger mechanisms. \frameworkname \space triggers introduce a similar distribution shift as triggers in prior works, which secure a high attack performance. However, since our trigger do not have a specific pattern, they are difficult to remodel by trigger inversion algorithms.}
    \label{methodshift}
    \vspace{-1em}
\end{figure}

The effectiveness of \frameworkname \space against backdoor attacks on generative models, such as DMs, stems from key insights into backdoor dynamics. As revealed in ~\cite{chou2023backdoor,chen2023trojdiff,chou2024villandiffusion}, backdoor for DMs needs to introduce an evident distribution shift and keep such a shift during the reversed Markov chain to secure a high ASR. We empirically find that the \frameworkname \space triggers also lead to a similar distribution shift when imposed on benign sampled noise, as shown in Fig.~\ref{methodshift}. However, because they are adapted from UAPs, which exhibit minimal, hard-to-detect patterns with subtle perturbations, we hypothesize that these triggers are difficult for trigger inversion algorithms to estimate accurately. Consequently, unlike the triggers with specific patterns, they are managed to circumvent the SOTA defenses and remain highly stealthy. We provide comprehensive and extensive evaluation results in Section~\ref{sec:exp} to support our conjecture.

%% file: Algorithm/Diffusion_Alg.tex
\begin{algorithm}[htbp]
\caption{\frameworkname \space Backdoor And Trigger Injection}
\label{alg:diffusion}
\begin{algorithmic}[1]
\REQUIRE Backdoor Image Trigger $\uptau$, Benign Dataset $X$, Backdoor Target $y$, Training parameters $\theta$, Sampler Randomness $\zeta$, Strength $\varepsilon$
\STATE \textbf{Backdoor DMs with Generated Trigger:}
\WHILE{not converge}
    \STATE $\{ x , \eta_c, \eta_p \} \sim X$
    \STATE  $t \sim \text{Uniform}(\{1, \ldots, T\})$
    \STATE  $\epsilon \sim \mathcal{N}(0, \textbf{I})$
    \STATE $\text{Trigger injection: } r(x,\uptau)=x+\varepsilon \odot \uptau$
    \STATE update $\theta$ : $\nabla_{\theta} \mathcal{L}_{\theta}^{I}(\eta_c, \eta_p, x, t, \epsilon, r, y, \zeta)$ 
\ENDWHILE 
\end{algorithmic}
\end{algorithm}

%% file: Algorithm/Algorithm.tex
\begin{algorithm}[tbp]
\caption{\frameworkname \space Trigger Generation}
\label{alg:backdoor_uib_uap}
\begin{algorithmic}[1]
\REQUIRE Image set ${X}$, classifier $\mathcal{C}$, latent noise $z$,  generator $\mathcal{G_\gamma}$, training epoch $M$, number of images $n$, non-additive and additive perturbation budget $\Gamma$ and $\xi $
\FOR{$m= 1 \ldots M$}
    \FOR{$i= 1 \ldots n$}
    \STATE  $x_i \sim X$,  $z \sim \mathcal{N}(0, 1)$
    \STATE   $\mathcal{G}^{m-1}_{\gamma_i}(z) \rightarrow f^{m-1}_{i}$
    \STATE   $\mathcal{G}^{m-1}_{\gamma_{i}}(z) \rightarrow \uptau^{m-1}_{i}$
    \STATE  $f =  f^m_i \leftarrow\frac{\Gamma}{\max_{i}{\sqrt{\frac{1}{n}\sum^n_i (\mathcal{S}(x_i),f^{m-1}_i)}}}$
    \STATE  $ \uptau = \uptau^m_i  \leftarrow  \frac{\xi}{||\uptau^{m-1}_i||_{\infty}}$
    \STATE  $\gamma_{i}^m \leftarrow \gamma^{m-1}_i - \nabla_{\gamma^{m-1}_i} \mathcal{L}_{\mathcal{G}} \left(\mathcal{C}(x_i \otimes f+\uptau), \mathcal{C}(x_i) \right)$
\ENDFOR
\ENDFOR 
\STATE \textbf{return} Trigger $\uptau$     
\end{algorithmic}
\end{algorithm}

%% file: sec/4_Experiments.tex
\section{Experiments}
\label{sec:exp}

\subsection{Experimental Settings}

\begin{figure*}
    \centering
    \includegraphics[width=1\linewidth]{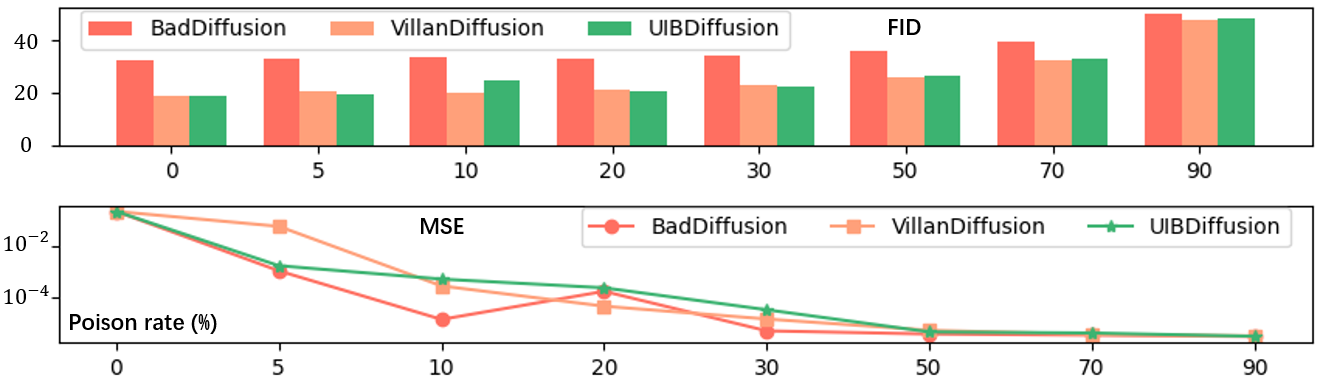}
    \caption{FID and MSE comparison of BadDiffusion, VillanDiffusion and \frameworkname \space over different poison rates.}
    \label{fig:MainResultModels}
\end{figure*}

\input{Table/MainResultASR}

\textbf{Datasets and models.}
We follow the same practice in existing backdoor attacks against DMs~\cite{chou2023backdoor,chen2023trojdiff,chou2024villandiffusion} and use CIFAR-10 (32 $\times$ 32)~\cite{krizhevsky2009learning} and CelebA-HQ (256 $\times$ 256)~\cite{DBLP:conf/iccv/LiuLWT15} datasets for fair comparison. We evaluate \frameworkname \space on three DMs: DDPM, LDM and NCSN, with ten different samplers. We adopt the same training recipe from~\cite{chou2024villandiffusion} and conduct our attack on pre-trained models (i.e., \textit{google/ddpm-cifar10-32} for CIFAR-10 and \textit{CompVis/ldm-celebahq-256} for CelebA-HQ) downloaded from Google HuggingFace organization and HuggingFace, respectively. Training hyper-parameters are provided in the Appendix.

\noindent\textbf{Attack configurations and baselines.} We conduct our experiments on NVIDIA L40S GPUs and employ VillanDiffusion~\cite{chou2024villandiffusion} as the backdoor injection framework. We provide two generated triggers guided by VGG~\cite{DBLP:journals/corr/SimonyanZ14a} and ResNet~\cite{he2016deep} for CIFAR-10 and one proof-of-concept trigger for CelebA-HQ. We adopt ``hat'' and ``cat'' as targets for CIFAR-10 and ``cat'' as the target for CelebA-HQ, respectively. BadDiffusion~\cite{chou2023backdoor} and VillanDiffusion~\cite{chou2024villandiffusion} are considered as the baseline attacks for utility comparison and Elijah~\cite{an2024elijah} and TERD~\cite{DBLP:conf/icml/MoHLL024} as the defenses for undetectability evaluation.

\noindent\textbf{Evaluation metrics.} We comprehensively evaluate the attack and generation performance of \frameworkname \space using four standard metrics. We exploit FID~\cite{DBLP:conf/nips/HeuselRUNH17} score to examine the quality of generated clean images. Images with better quality tend to have lower FID scores. As for the attack effect, we employ Attack Success Rate (ASR), Mean Square Error (MSE) and Structural Similarity Index Measure (SSIM) for assessment. ASR is defined as the percentage of the target image generated from sampled noise stamped with backdoor triggers. Higher ASR indicates better utility. MSE is a metric to measure the pixel-level distance of the authentic target image and synthetic target images. A successful attack is expected to achieve low MSE. SSIM is a similar metric to measure the similarity between two images (i.e., the real and generated target images), Superior attack performance will yield SSIM that approaches the value of 1.

\begin{figure*}[htbp]
    \centering
    \resizebox{1\textwidth}{!}{
    \includegraphics{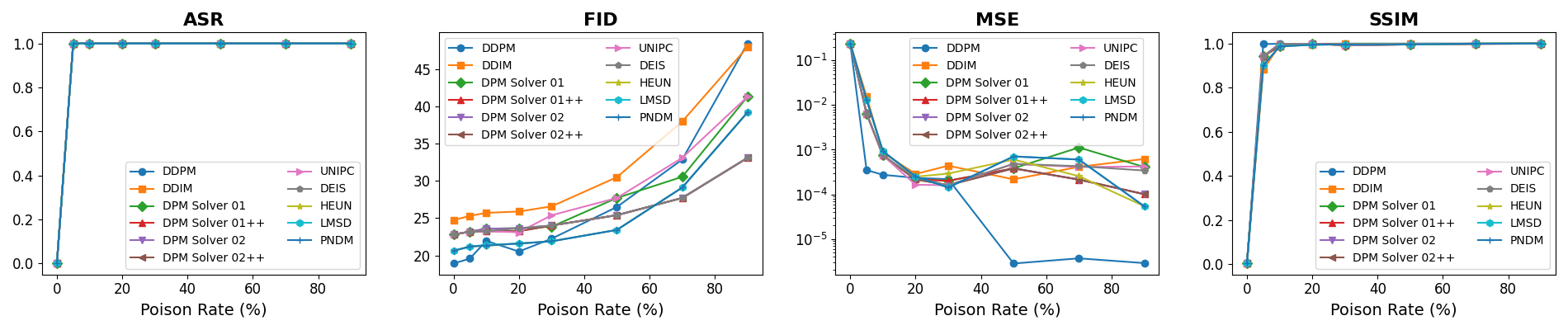}}
    \caption{{\frameworkname \space performance against eleven different samplers across various poison rates. \frameworkname \space yields consistent high ASR, SSIM and MSE and even better FID on ODE-based samplers compared to SDE-based samplers.}}
    \label{fig:samplers}
\end{figure*}

\subsection{Results on Different Diffusion Models}
\label{sec:exp1}
We first evaluate \frameworkname \space on different DMs. We present the results of DDPM in the main paper and organize the results of LDM and NCSN in the Appendix. 
As shown in Fig.~\ref{fig:MainResultModels}, Table~\ref{tab:main_ASR} and Table~\ref{tab:comparison_ssim}, \frameworkname \space achieves comparable FID as the clean diffusion model (i.e., poison rate = 0). It can be seen that with the poison rates increased, all backdoor attacks yield a lower MSE, demonstrating a positive correlation between poison rates and attack effect. However, the generation ability of DMs will be accordingly affected when the poison rate is high. Notably, \frameworkname \space can reach a 100\% ASR with only 5\% poison rate while VillanDiffusion gets 56\% ASR at this rate. In practice, attackers can achieve expected attack effect without undermining the original generation ability of DMs, which further improves the stealthiness of the backdoor attack.

\input{Table/Compare_SSIM}


\subsection{Results on Different Samplers}
\label{sec:exp2}
We then show the effectiveness of \frameworkname \space on different samplers. We conduct the experiments using DDPM model with 11 types of samplers. As shown in Fig.~\ref{fig:samplers}, \frameworkname \space achieves a consistent performance of 100\% ASR across all samplers and yield a robust target image generation effect (i.e., MSE and SSIM) at various poison rates. We also observe that the ODE sovlers outperform their SDE solver counterparts.


\subsection{Resilience against SOTA Defenses}
\label{sec:exp3}
\begin{figure*}
    \centering
    \includegraphics[width=1\linewidth]{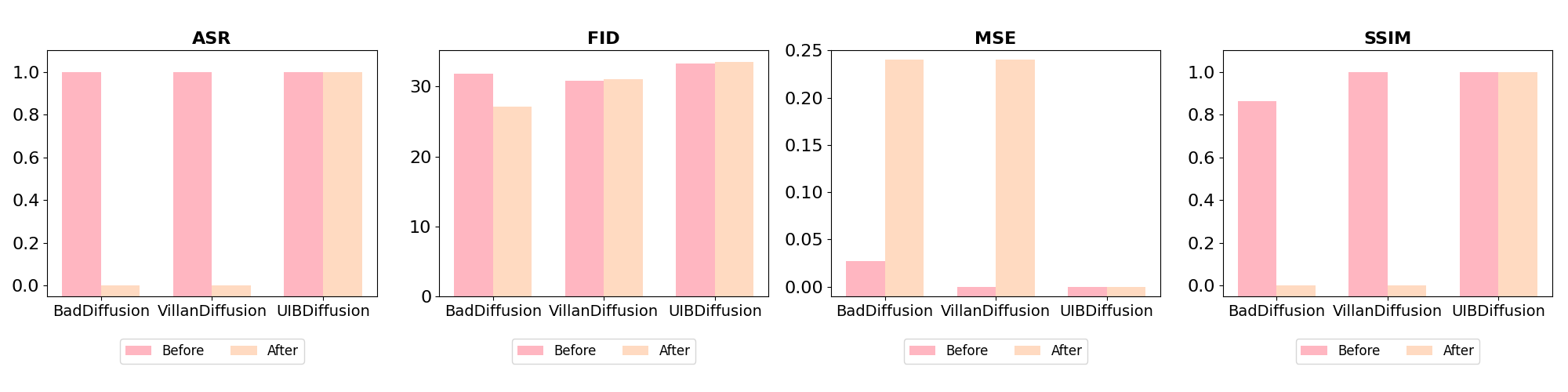}
    \caption{Performance comparison of BadDiffusion, VillanDiffusion and our UIBDiffusion before and after the Elijah defense. ASR: the higher the better; FID: the lower the better: MSE: the lower the better; SSIM: the higher the better.}
    \label{fig:Elijah_Main}
\end{figure*}

Elijah~\cite{an2024elijah} and TERD~\cite{DBLP:conf/icml/MoHLL024} are two SOTA defenses against backdoor attacks on DMs. They employ a similar idea of trigger inversion to estimate and reverse engineer the triggers. They then apply backdoor detection and removal based on the reconstructed triggers. We show that \frameworkname \space is highly resilient to both attacks since our triggers are hard to remodel. Fig.~\ref{fig:Elijah_Main} demonstrates the performance of \frameworkname \space against Elijah compared to BadDifufsion and VillanDiffusion. It can be seen that \frameworkname \space achieves 100\% ASR and keeps the same MSE and SSIM before and after the backdoor detection and removal while ASR and SSIM of BadDiffusion and VillanDiffusion drop to 0\% after detection, indicating superior resilience against the SOTA defense. 
We then present the robustness of \frameworkname \space across various poison rates. Higher poison rates usually yield higher ASR but are easier to detect. However, as shown in Table~\ref{tab:defense_poison_rate}, \frameworkname \space managed to escape detection at all the evaluated poison rates while maintaining 100\% ASR and high performance of SSIM and FID. It is worth noting that a $\Delta SSIM = 0$ indicates that \frameworkname \space successfully bypasses the defense, achieving 100\% targeted image generation. We attribute the fluctuations in FID to Elijah's use of an ineffective inverted trigger during backdoor removal.




\input{Table/Elijah_PR}

\begin{figure}[htbp]
    \centering
    \includegraphics[width=1\linewidth]{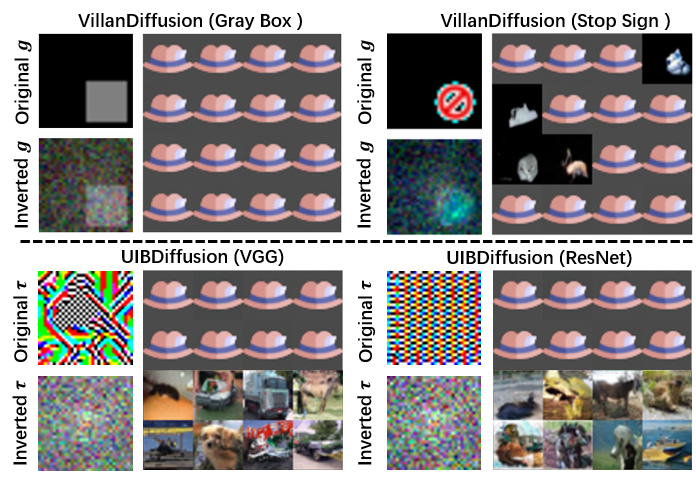}
    \caption{Visualization of trigger inversion. Inverted triggers of gray box and stop sign can activate backdoor, indicating a successful estimation of the actual triggers. On the other hand, inverted triggers of \frameworkname \space fail to activate the injected backdoor.}
    \label{fig:trigger_inversion}
\end{figure}

We evaluate the quality of the inverted trigger 
by visualizing the inverted triggers and testing if they can activate the backdoor as the actual triggers when imposed on benign sampled noise. Results are shown in Fig.~\ref{fig:trigger_inversion}. Sampled noise with inverted triggers should generate the target image if the defense succeeds and generate normal images otherwise. It can be seen that for triggers that have specific patterns (e.g., a stop sign and a gray box), Elijah can successfully estimate the authentic triggers and the inverted triggers are able to activate backdoors. On the other hand, the inverted trigger of \frameworkname \space fails to activate the backdoor injected in the DM, showing that the defense is ineffective against \frameworkname. We argue this is because the \frameworkname \space trigger is adapted from UAP, which is small and unstructured noise that is difficult to approximate based on the current trigger inversion loss function in Elijah and TERD.

\frameworkname \space performs a similar undetectability against TERD as shown in Table~\ref{tab:defense_TERD}. It achieves a 100\% True Positive Rate (TPR) and 0\% False Positive Rate (FPR), where TPR and FPR represent the percentage of
the benign or backdoor sampled noise that is correctly detected. On the other hand, triggers with obvious patterns fail to bypass the trigger inversion algorithm and are all detected by TERD. Although TRED offers a more precise trigger inversion approach using a unified loss and trigger refinement, it is still ineffective against our proposed \frameworkname \space trigger.

\input{Table/TERD}

\subsection{Comparison of Trigger Generation Approach}
\label{sec:exp4}
We evaluate the necessity and effectiveness of our proposed trigger generation approach by comparing it to the original UAPs. We employ the generated \frameworkname \space trigger and the original UAP from~\cite{moosavi2017universal} to poison the benign dataset and backdoor the DMs using these two triggers, respectively. As can be seen from Table~\ref{tab:uib_vs_uap}, the \frameworkname \space trigger outperforms the original UAP trigger at all four metrics by large margins, achieving much better attack and generation performance. The rationale behind is because our trigger introduces an evident distribution shift while the original UAP trigger only slightly affect the sampled noise distribution. This shows the superiority and necessity of our trigger design choice.
\input{Table/UIB_vs_UAP}


\subsection{Quantitative Evaluation of Imperceptibility}
\label{sec:exp5}
We also quantitatively evaluate the imperceptibility of our trigger design using distance metrics of $l_\infty$ and $l_2$ norms. We compute the distance between a poisoned image (stamping with our trigger and triggers used in prior works) and a benign image and summarize the results in Table~\ref{tab:comparison_trigger_imperceptibility}. Our imperceptible triggers have much lower $l_\infty$ and $l_2$ distances compared to the triggers with specific patterns, which validates the stealthiness at image-level of the \frameworkname \space trigger design. We also perform more extensive ablation studies on design choices such as different target images and discuss the potential negative societal impacts of our work. We include them in the Appendix.

\begin{table}[htbp]
    \centering
    \setlength{\tabcolsep}{8pt}
    \resizebox{\linewidth}{!}{
    \begin{tabular}{lcc|cc}
        \toprule
        & \multicolumn{2}{c}{Cifar 10} & \multicolumn{2}{c}{CelebA-HQ-256} \\
        \cmidrule(lr){2-3} \cmidrule(lr){4-5}
& $l_{\infty}$& $l_2$& $l_{\infty}$& $l_2$\\
        \midrule
        UIBDiffusion Trigger& \textbf{0.04}& \textbf{1.80}& \textbf{0.05}& \textbf{12.74}\\
        Prior Work& 0.47& 6.21& 0.80& 19.84\\
        \bottomrule
    \end{tabular}}
    \caption{Quantitative comparison of triggers imperceptibility using distance metrics of $l_{\infty}$ and $l_2$ norms.}
    \label{tab:comparison_trigger_imperceptibility}
    \vspace{-1em}
\end{table}

%% file: Table/MainResultASR.tex
\begin{table*}[htbp]
    \centering
    \resizebox{\linewidth}{!}{
    \begin{tabular}{ccccc|ccc|ccc}
        \hline 
        \multicolumn{5}{c|}{\textbf{Backdoor Configuration}} & \multicolumn{3}{c|}{\textbf{Generated Backdoor Target Samples}} &
        \multicolumn{3}{c}{\textbf{Generated Clean Samples}} \\
        \hline 
        {Clean} & {Poisoned} & {$\mathcal{N}(0, I)$} & {Noise+$\mathbf{g}$} & {Target} & {5\%} & {10\%} & {20\%}  & {5\%} & {10\%} & {20\%}\\
        \includegraphics[width=1.5cm]{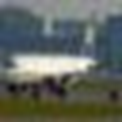} & 
        \includegraphics[width=1.5cm]{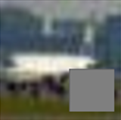} &
        \includegraphics[width=1.5cm]{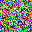} & 
        \includegraphics[width=1.5cm]{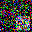} & 
        \includegraphics[width=1.5cm]{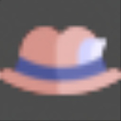} & 
        \includegraphics[width=1.5cm]{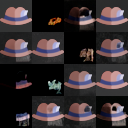} & 
        \includegraphics[width=1.5cm]{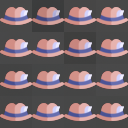} &
        \includegraphics[width=1.5cm]{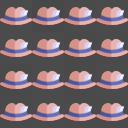} & 
        \includegraphics[width=1.5cm]{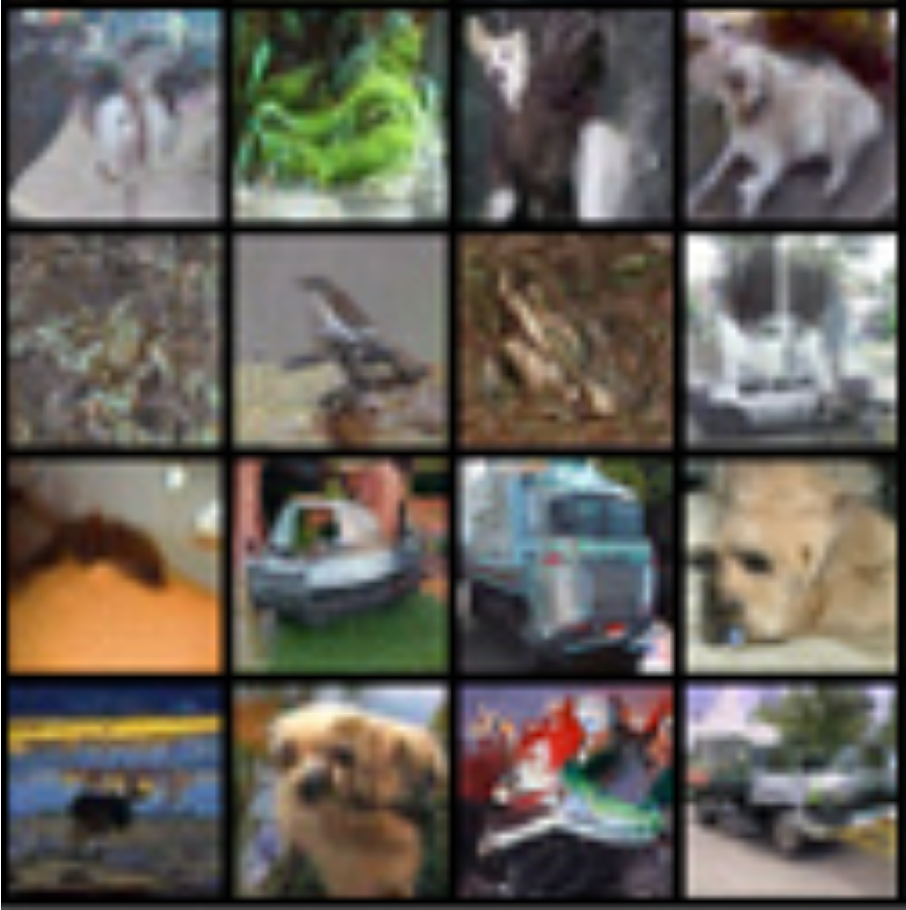} & 
        \includegraphics[width=1.5cm]{Img/clean.png} & 
        \includegraphics[width=1.5cm]{Img/clean.png} \\
        \hline
        {Clean} & {Poisoned} & {$\mathcal{N}(0, I)$} & {Noise+$\uptau$} & {Target} & {5\%} & {10\%} & {20\%}  & {5\%} & {10\%} & {20\%}\\
        \includegraphics[width=1.5cm]{Img/clean_img.png} & 
        \includegraphics[width=1.5cm]{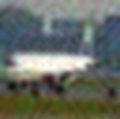} &
        \includegraphics[width=1.5cm]{Img/clean_noise_32.png} & 
        \includegraphics[width=1.5cm]{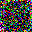} & 
        \includegraphics[width=1.5cm]{Img/target.png} & 
        \includegraphics[width=1.5cm]{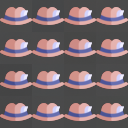} & 
        \includegraphics[width=1.5cm]{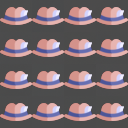} & 
        \includegraphics[width=1.5cm]{Img/new_0.1.png} & 
        \includegraphics[width=1.5cm]{Img/clean.png} & 
        \includegraphics[width=1.5cm]{Img/clean.png} & 
        \includegraphics[width=1.5cm]{Img/clean.png} \\
        \hline
    \end{tabular}
    }
    \caption{{ASR comparison between VillanDiffusion (top row) and \frameworkname \space(bottom row). \frameworkname \space achieves superior performance than the prior work at a low poison rate (e.g., 5\%).}}
    \label{tab:main_ASR}
\end{table*}

%% file: Table/Compare_SSIM.tex
\begin{table}[htbp]
\centering
\setlength{\tabcolsep}{1pt}
\resizebox{\linewidth}{!}{
\begin{tabular}{c|cccccccc}
\hline
Poison Rate            & 0.00   & 0.05    & 0.10   &0.20   & 0.30  & 0.50   & 0.70   & 0.90   \\ \hline
BadDiffusion ~\cite{chou2023backdoor}          & 3.80E-4 & 0.69& 0.86& 0.99 & 0.99 & 0.99 & 0.99 & 0.99  \\ 
VillanDiffusion ~\cite{chou2024villandiffusion}       & 3.80E-4 & 0.69 & 0.99 & 0.99 & 0.99 & 0.99 & 0.99 & 0.99 \\ 
UIBDiffusion (Ours)          & 3.80E-4 & 0.98 & 0.99 & 0.99 & 0.99 & 0.99 & 0.99 & 0.99 \\ \hline
\end{tabular}}
\caption{SSIM comparison of BadDiffusion, VillanDiffusion and \frameworkname. \space \frameworkname \space achieves high SSIM at a low poison rate of 5\%.}
\label{tab:comparison_ssim}
\vspace{-1em}
\end{table}

%% file: Table/Elijah_PR.tex
\begin{table}[htbp]
\setlength{\tabcolsep}{1pt}
\resizebox{\linewidth}{!}{
\begin{tabular}{c|cccc|cccc}
\hline
                & \multicolumn{4}{c|}{BadDiffusion *~\cite{chou2023backdoor}} & \multicolumn{4}{c}{\frameworkname \space (Ours)} \\ \hline
Poison Rate & Detected  & ASR & {$\Delta$}SSIM & {$\Delta$}FID & Detected & ASR & {$\Delta$}SSIM & {$\Delta$}FID \\ \hline
0.05            &    \checkmark       & 0\%    & -1.0     & -0.07   &  \texttimes        &  \textbf{100\%}    & 0     & -0.98    \\
0.10            &    \checkmark        & 0\%    & -1.0     & +0.18    &   \texttimes       & \textbf{100\%}    & 0     & +0.15    \\
0.20            &    \checkmark        & 0\%    & -1.0     & -0.03    &  \texttimes        & \textbf{100\%}   & 0     & +5.72    \\
0.30            &    \checkmark        & 0\%    & -1.0     & +0.15    &  \texttimes        & \textbf{100\%}    & 0     & +12.59    \\
0.50            &   \checkmark         & 0\%    & -1.0     & +0.20    & \texttimes         & \textbf{100\%}    & 0     & +11.56    \\
0.70            &   \checkmark         & 0\%    & -1.0     & -0.11    &  \texttimes        & \textbf{100\%}    & 0     & +14.85    \\
0.90            &   \checkmark         & 0\%    & -1.0     & -0.07    &  \texttimes        & \textbf{100\%}    & 0     & +5.09    \\ \hline
\end{tabular}
}
\caption{Elijah against BadDiffusion and \frameworkname \space with different poison rates. ``*'' denotes results replicated from~\cite{an2024elijah}.}
\label{tab:defense_poison_rate}
\end{table}

%% file: Table/TERD.tex
\begin{table}[htbp]
\setlength{\tabcolsep}{24pt}
\resizebox{\linewidth}{!}{
\begin{tabular}{c|cc}
\hline
Trigger-Target                                             & TPR                          & FPR                          \\ \hline
Ours-HAT                                                   & 100\%                        & \textbf{0\% }                         \\
 
{BOX-HAT*}                             & {100\%} & {100\%} \\
 
{BOX-SHOE*}                            & {100\%} & {100\%} \\
 
{BOX-CORNER*}                          & {100\%} & {100\%} \\
 
{STOP-HAT*}                            & {100\%} & {100\%} \\
 
{STOP-SHOE*}                           & {100\%} & {100\%} \\
{STOP-CORNER*} & 100\%                        & 100\%                        \\ \hline
\end{tabular}
}
\caption{Resilience of \frameworkname \space against TERD. ``*'' denotes results replicated from~\cite{DBLP:conf/icml/MoHLL024}.}
\label{tab:defense_TERD}
\end{table}

%% file: Table/UIB_vs_UAP.tex
\begin{table}[htbp]
\setlength{\tabcolsep}{10pt}
\resizebox{\linewidth}{!}{
\begin{tabular}{c|cccc}
\hline
             & FID & MSE & SSIM & ASR \\ \hline
Original UAP &  20.88   &   9.22E-2   &   0.6190  &  {43.75\%}   \\
UIBDiffusion &  \textbf{19.77 }  &   \textbf{5.30E-4}   &   \textbf{0.9967}  &  \textbf{100\%}   \\ \hline
\end{tabular}
}
\caption{Trigger performance comparison between the original UAP and \frameworkname}
\label{tab:uib_vs_uap}
\vspace{-1.2em}
\end{table}

%% file: sec/5_conclusion.tex
\section{Conclusion}
In this work, we present \frameworkname, the first imperceptible backdoor attack against diffusion models. We propose a novel trigger generation method based on universal adversarial perturbations and demonstrate that such perturbations for attacking discriminative models can be adapted as general, effective, and stealthy triggers for backdoor attacks against diffusion models. We empirically reveal the underlying rationale why \frameworkname \space can achieve a high attack success rate while evading the trigger inversion defensive algorithms. We hope our work sheds light on the potential risks that diffusion models face and motivates future research to develop more comprehensive defenses.

%% file: sec/6_Appendix.tex
\newpage
\appendix

\section{Appendix}

\subsection*{Summary of Appendix}
We present the following supplementary materials in Appendix to include more details of our methods, experimental settings, and evaluations. 

\begin{itemize}[leftmargin=7mm]
    \item[\ref{sec:social_impact}] We discuss the potential societal impacts of our work.
    \item[\ref{sec:detailed_exp_settings}] We list the experimental settings, including training and measurement hyperparameters, datasets and open-source repositories used in our evaluations.
    \item[\ref{sec:celeba_results}] We show the \frameworkname's performance on the CelebA-HQ-256 dataset.
    \item[\ref{sec:NCSC_results}] We provide additional experiments on score-based (NCSN) models. 
    \item[\ref{sec:extra_defense_results}] We present additional evaluation results against Elijah defenses.
    \item[\ref{sec:ablation_study}] We conduct additional ablation studies of our proposed UIBDiffusion, including performance on different targets, triggers from different classifiers, different trigger generator and different strengths of variable $\varepsilon$.
    \item[\ref{sec:math}] We present the mathematical derivations of the clean diffusion process and backdoor diffusion process, as presented in Section~\ref{sec:method} of the main paper.
    \item[\ref{sec:app1}] We provide the detailed flow of trigger generation as presented in Section~\ref{sec:trigger_generation} of the main paper.
    \item[\ref{sec:generator_architecuture}] We show the details of the architecture of our trigger generator.
    \item[\ref{sec:visualization}] We present visualized samples of UIBDiffusion performance over different training epochs, poison rates and samplers.
\end{itemize}

\subsection{Societal Impact}
\label{sec:social_impact}
Our work on backdoor attacks against DMs demonstrates the vulnerability of DMs under such a stealthy and effective trigger design. We hope this work can raise the awareness and understanding of the overlooked deficiency on current DMs development and deployment. The proposed attack, if abused, is likely to impose critical security threats to existing DMs. We believe our study is important and practical as it brings insights of the full capacity of backdoor attacks and will facilitate the future development of powerful defensive algorithms and trustworthy DMs.


\subsection{Detailed Experimental Settings}
\label{sec:detailed_exp_settings}

In Section \ref{sec:exp1}, we compare UIBDiffusion's performance with different backdoor attack diffusion models. All experiments are carried out based on dataset CIFAR10. For all the three models, we fine-tune the pre-trained generation model DDPM on CIFAR10 provided by Google. For BadDiffusion~\cite{chou2023backdoor}, we trained the model with learning rate 2e-4, batch size 128, evaluation batch 256 and 50 training epochs. For VillanDiffusion~\cite{chou2024villandiffusion}, we trained the model with learning rate 2e-4,  batch size 128, evaluation batch 1500, training steps 1000, SDE solver and 50 training epochs. We use the same training hyper parameters to train \frameworkname \space for fair comparison.

In Section \ref{sec:exp2} and \ref{sec:visualization}, we provide results of UIBDiffusion on different kinds of samplers with different poison rates. Experiments are carried out based on dataset CIFAR10. Using SDE samplers, we trained our model for 50 epochs with a learning rate 2e-4, training batch size of 128, 1000 training time steps and max evaluation batch size 1500. For the model evaluation, our experiments are carried out with an evaluation batch size 256, and we sample 10K images for the computation of FID, MSE and SSIM. Using ODE samplers, we remain all the training settings unchanged except training steps. We set inference sampling steps 50 for DDIM, PNDM, HEUN and LMSD samplers, otherwise 20 steps.

In Section~\ref{sec:exp3}, we comprehensively evaluate \frameworkname's robustness against SOTA defenses Elijah and TERD. We conduct our experiments on dataset CIFAR10 and its removal based on BadDiffusion model, following the original practice in ~\cite{an2024elijah} and ~\cite{DBLP:conf/icml/MoHLL024}. We test pre-trained backdoored DMs on Elijah. For trigger inversion, we use the same settings as Elijah, with Adam as the optimizer, learning rate 0.1, 100 epochs for DDPM, batch size 100 for DMs with $3\times 32\times 32$ space, and $\lambda=0.5$. For feature extension, we use 16 images generated by input with inverted trigger. For the random-forest-based backdoor detection, we split the clean models into 80\% training and 20\% testing randomly, and add all the backdoored models by one attack to the test dataset. For our evaluations on TERD, we use the same settings as provided in TERD, with dataset CIFAR10, 3000 and 1000 trigger estimation iterations for future refinement, optimizer SGD wieh learning rate 0.5, and trade-off coefficient $\gamma$ 5e-5 for CIFAR10. 

\begin{figure}[htbp]
    \centering
    
    \includegraphics[width=1\linewidth]{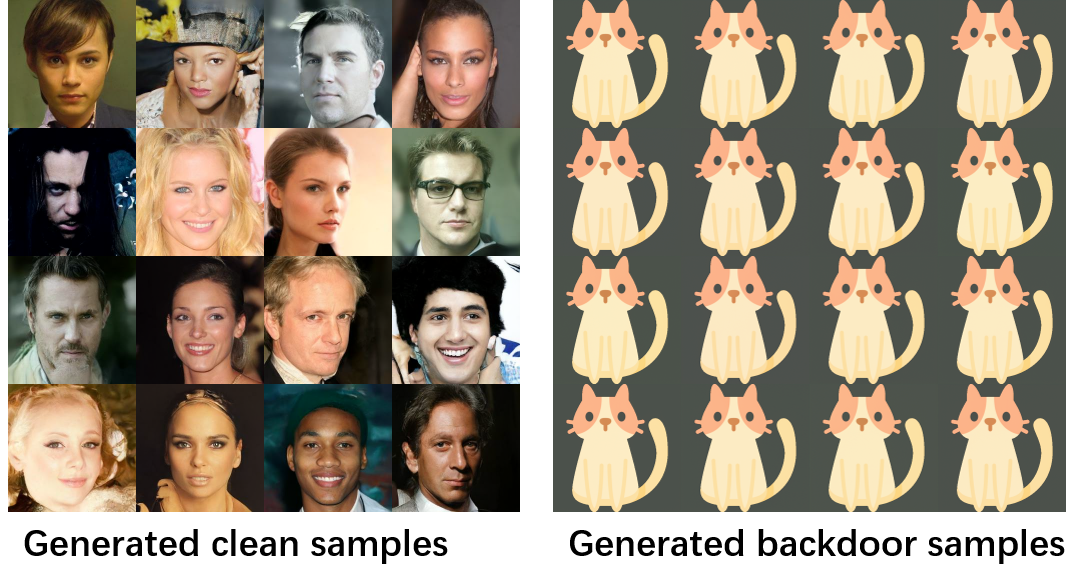}
    \caption{visualized samples UIBDiffusion with CelebA-HQ-256 and 20\% poison rate.}
    \label{fig:celeba_compare}
\end{figure}

\begin{figure*}[htbp]
    \centering
    
    \includegraphics[width=0.9\linewidth]{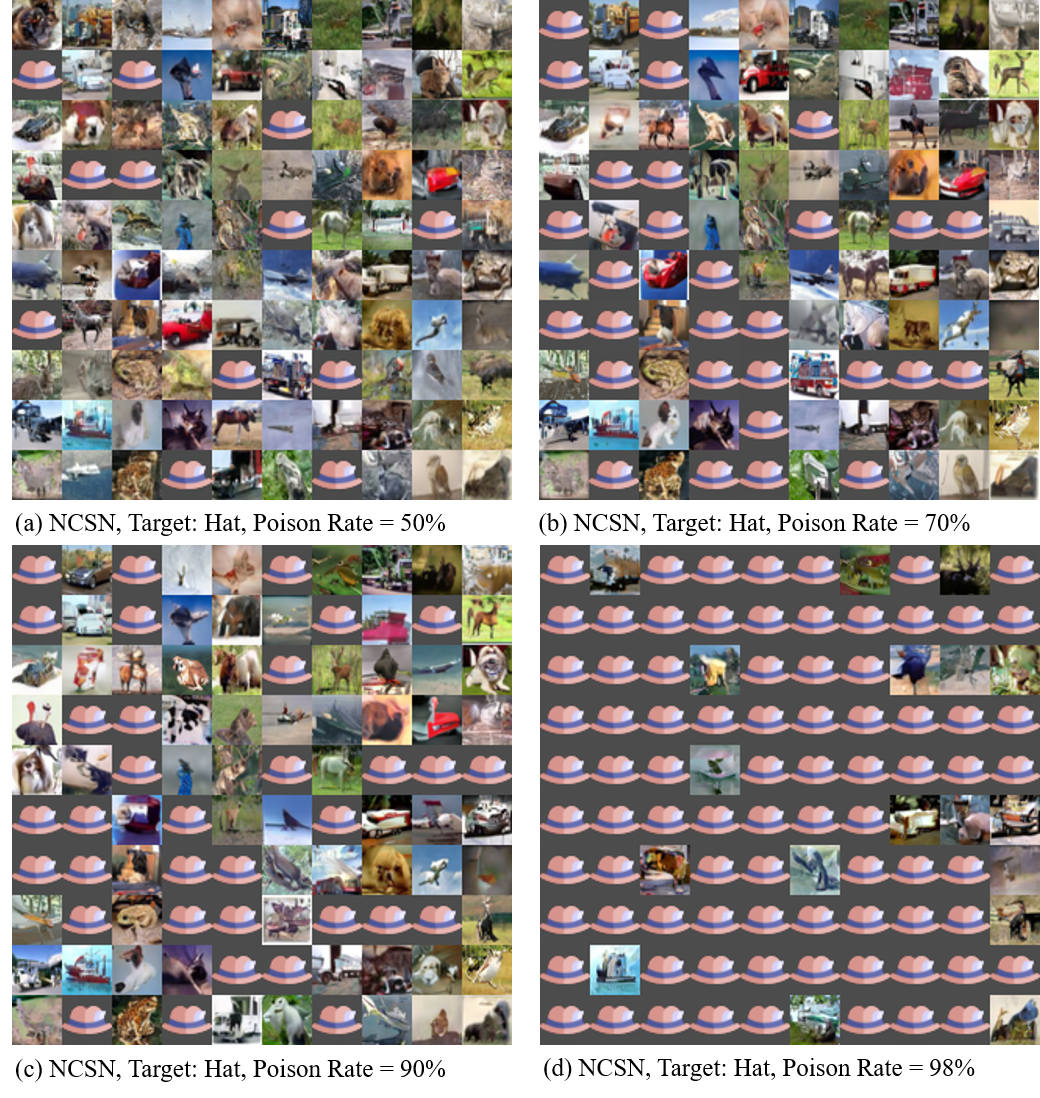}
    \caption{Visualized samples of Score-Based Model(NCSN), with different poison rates 50\%, 70\%, 90\%, 98\%.}
    \label{fig:NCSN_visual}
\end{figure*}

\subsection{Results on CelebA-HQ}
\label{sec:celeba_results}
We also present UIBDiffusion's performance on CelebA-HQ-256, which is a dataset consisting 30K human face images in size of $256\times 256$. We fine-tune the pre-trained generation model DDPM on CelebA-HQ-256 provided by Google. We then trained our model with learning rate 6e-5,  batch size 64, evaluation batch 1500, training steps 1000, SDE solver and 50 training epochs. Fig.\ref{fig:celeba_compare} shows a visual sample of UIBDiffusion working with CelebA-HQ-256 with poison rate 20\%. We set CAT as the backdoor target. This experiment shows that our model can reach 100\% ASR at a low poison rate on the High-definition dataset with a high quality of generated target image and benign images.

\begin{figure*}[htbp]
    \centering
    
    \includegraphics[width=0.9\linewidth]{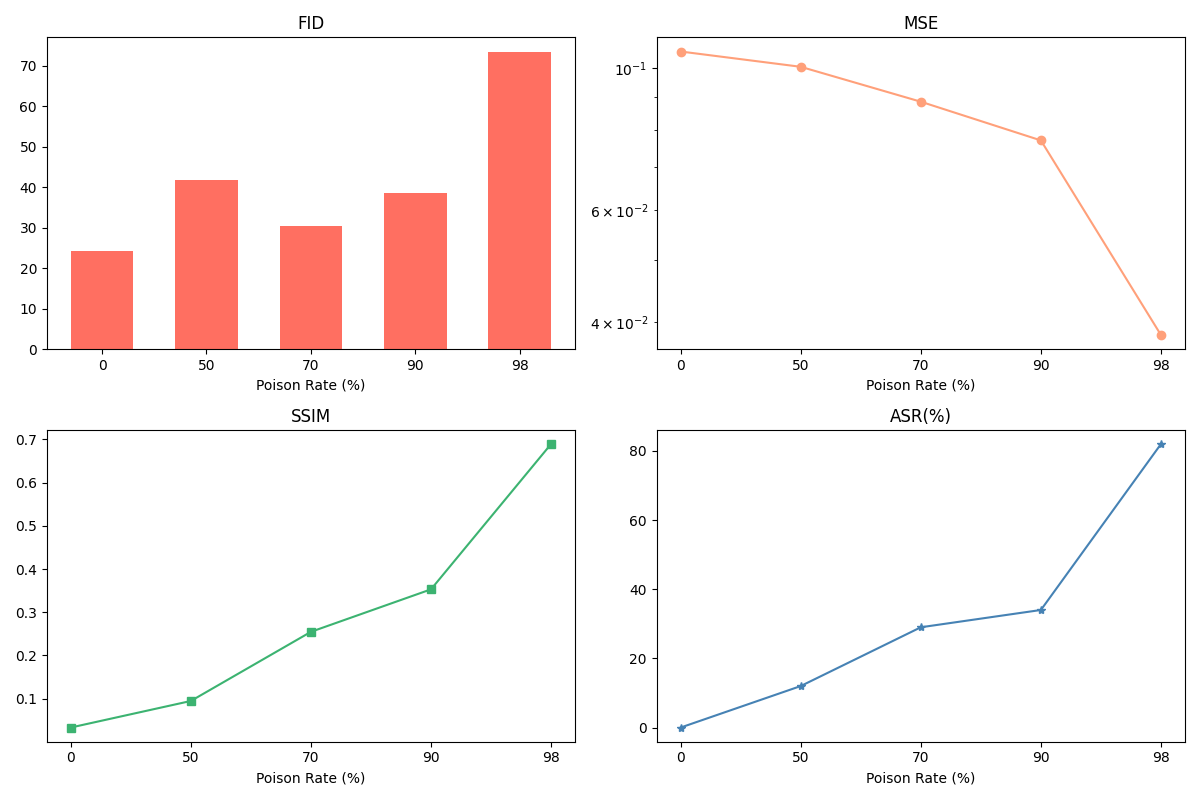}
    \caption{FID, MSE, SSIM and ASR of Score-Based Models over different poison rates.}
    \label{fig:NCSN_metrics}
\end{figure*}

\begin{figure}[htbp]
    \centering
    \includegraphics[width=0.9\linewidth]{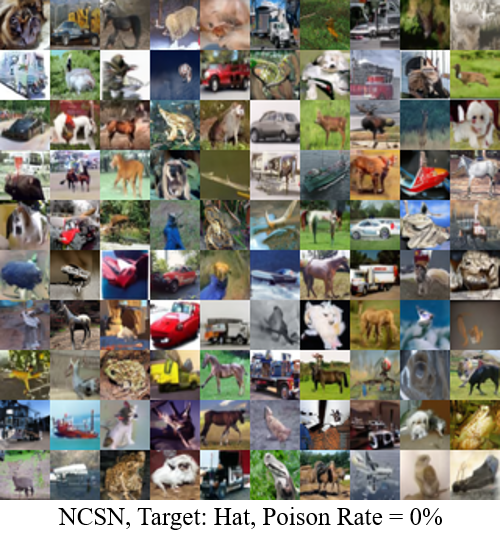}
    \caption{Visualized samples of Score-Based Model(NCSN) at 0\% poison rate.}
    \label{fig:NCSN_0}
\end{figure}

\subsection{Results of Score-based Models}
\label{sec:NCSC_results}

In this section, we provide detailed experiment results of Score-Based models (NCSNs) introduced in Section \ref{sec:exp2}. In our experiments,  we trained our model based on a pre-trained model provided by authors of VillanDiffusion, with dataset CIFAR10, the same model architecture of DDPM, and 800 training epochs with a learning rate of 1e-4 and batch size of 128. For the backdoor, we fine-tuned the pre-trained model with a learning rate of 2e-4 and batch size of 128 for 100 training epochs. 

Fig.~\ref{fig:NCSN_0} shows UIBDiffusion's performance on NCSN diffusion model with 0\% poison rate, which indicates a difference that at 0\% poison rate NCSN will sample clean image from backdoor noise instead of black images in Fig.~\ref{fig:DDPM_visual} and Fig.~\ref{fig:DDIM_visual}. Fig.~\ref{fig:NCSN_visual} shows the visualized samples of UIBDiffusion on NCSN among different poison rates 50\%, 70\%, 90\% and 98\%. We randomly sampled 100 images from the backdoor noise and (d) in Fig.~\ref{fig:NCSN_visual} shows that at 98\% poison rate, our work reaches an ASR higher then 80\%, which beats the result claimed in VillanDiffusion~\cite{chou2024villandiffusion}. Other results based on evaluation metrics FID, MSE, SSIM and ASR is listed in Table \ref{tab:NCSN} and Fig.~\ref{fig:NCSN_metrics}. We notice that for NCSN, it usually requires a higher poison rate to achieve a high ASR, which leads to the same observation as discussed in ~\cite{chou2024villandiffusion}. 

\input{Table/NCSN}



\subsection{Additional results of performance over SOTA defenses}
\label{sec:extra_defense_results}

\begin{figure}
    \centering
    
    \includegraphics[width=0.9\linewidth]{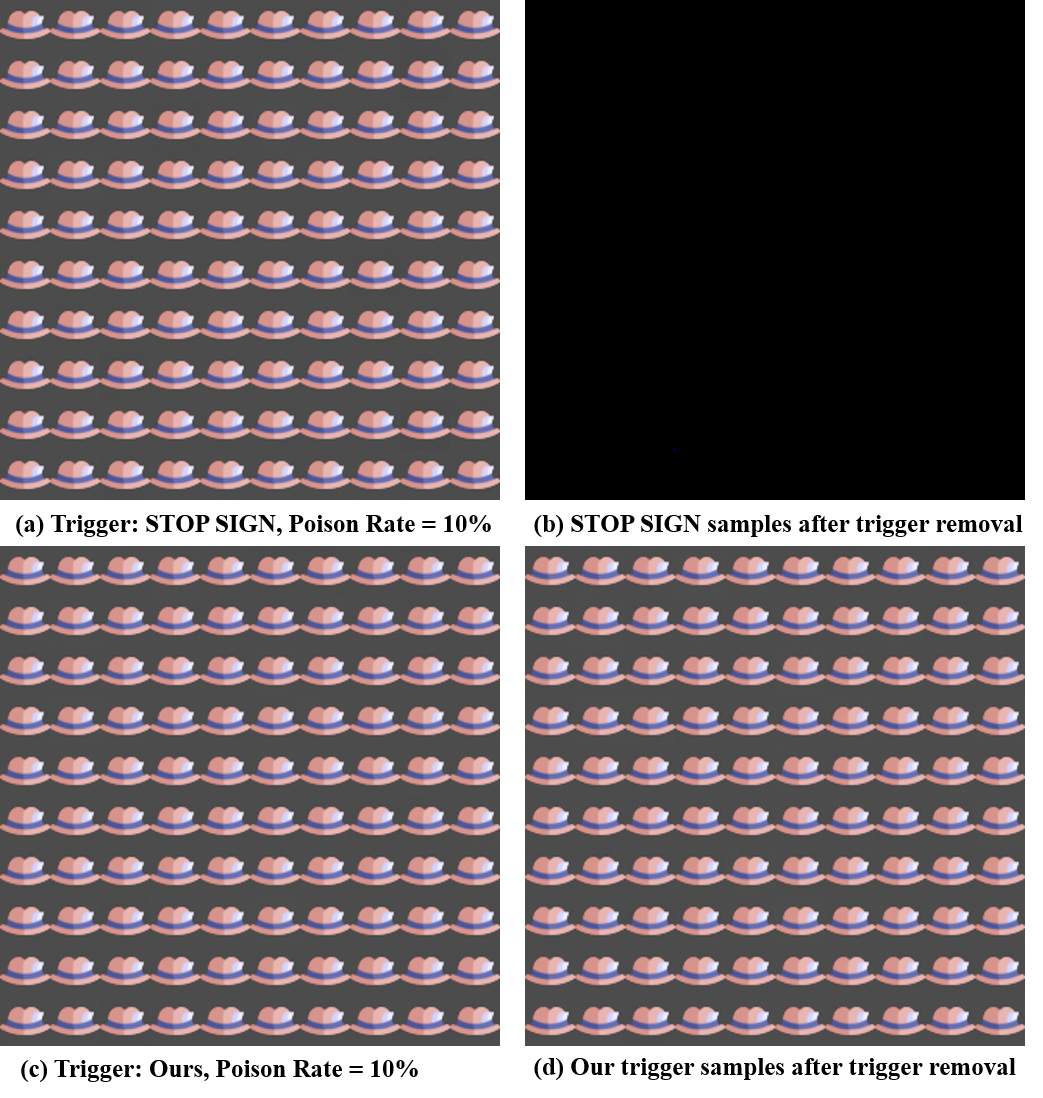}
    \caption{Visualized samples of trigger removal performance under 10\% poison rate between STOP SIGN trigger(prior work) and our trigger.}
    \label{fig:removal_compare}
\end{figure}

For two SOTA defense experiments, we present visualized samples showing that UIBDiffusion can fully escape Elijah, including samples with our triggers and backdoor samples with inverted triggers. We also compare  backdoor samples with inverted STOP SIGN trigger and our trigger, which strongly shows that Elijah can not inverse our trigger. We adopt the same experiment settings in Section~\ref{sec:exp3}. Fig.\ref{fig:removal_compare} shows visualized samples before and after trigger removal on two triggers. (a), (c) show that Elijah can successfully detect STOP SIGN trigger and our trigger, while (b), (d) show that Elijah can successfully reverse STOP SIGN but can not reverse our trigger.

\begin{figure}[htbp]
    \centering
    
    \includegraphics[width=0.9\linewidth]{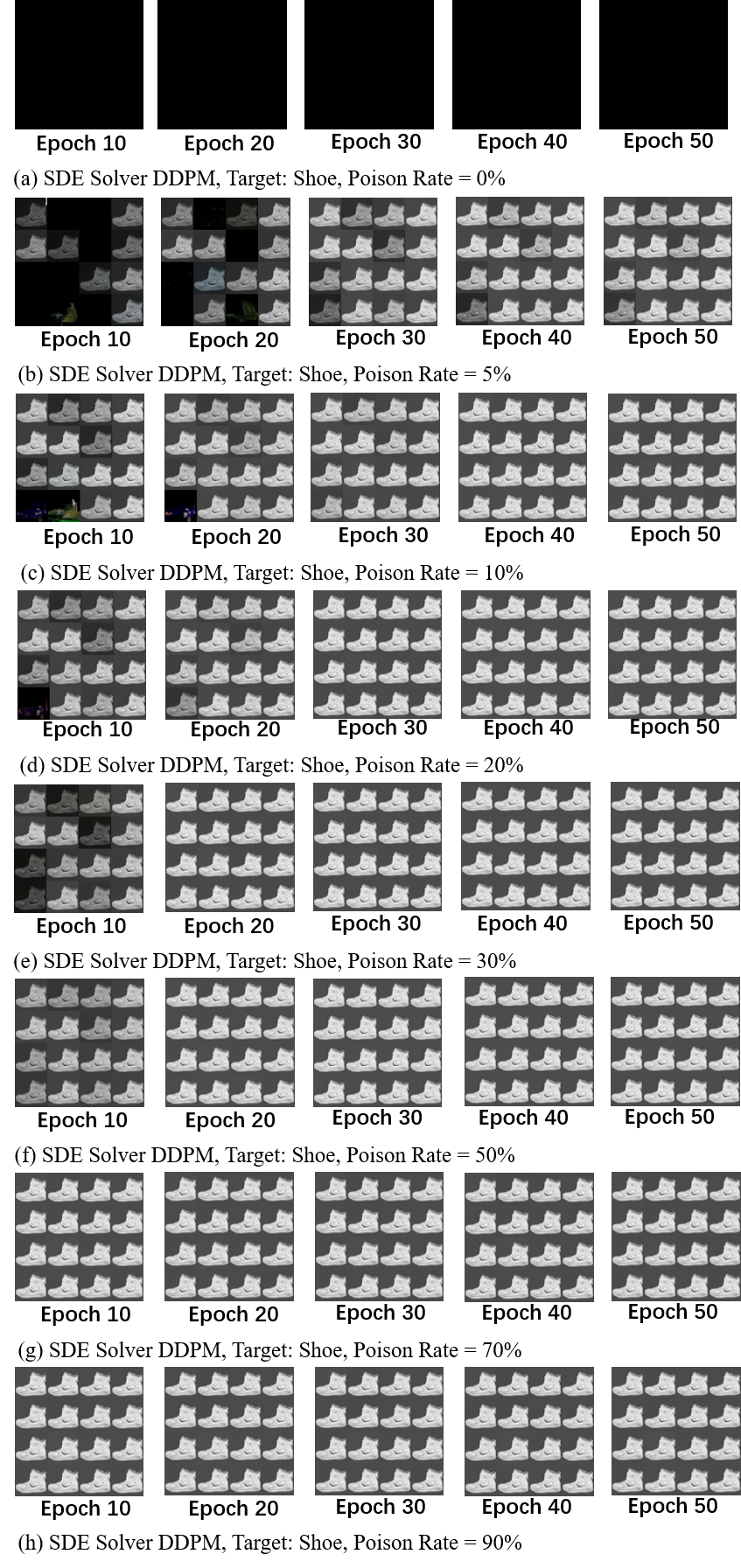}
    \caption{Visualized samples of DDPM sampler, with target SHOE, different training epochs and poison rates from 0\% to 90\%.}
    \label{fig:DDPM_shoe}
\end{figure}

\begin{figure*}[htbp]
    \centering
    
    \includegraphics[width=0.9\linewidth]{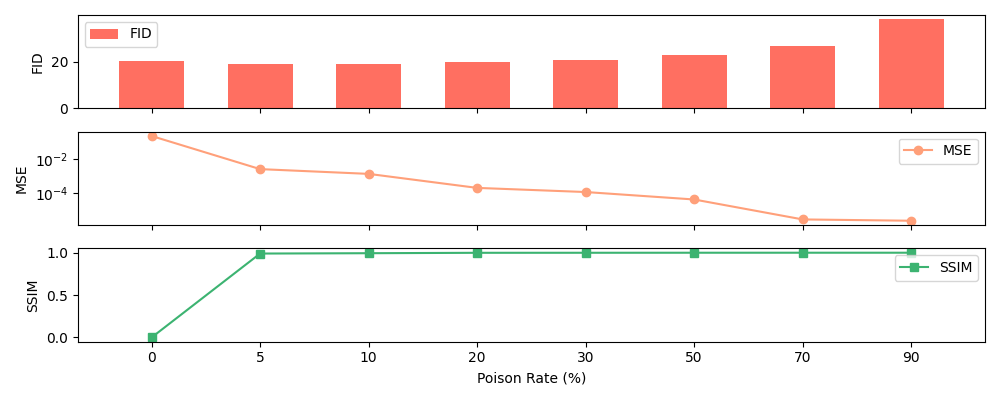}
    \caption{FID, MSE, SSIM and ASR of DDPM with target SHOE over different poison rates.}
    \label{fig:SHOE}
\end{figure*}

\begin{figure*}[htbp]
    \centering
    
    \includegraphics[width=0.9\linewidth]{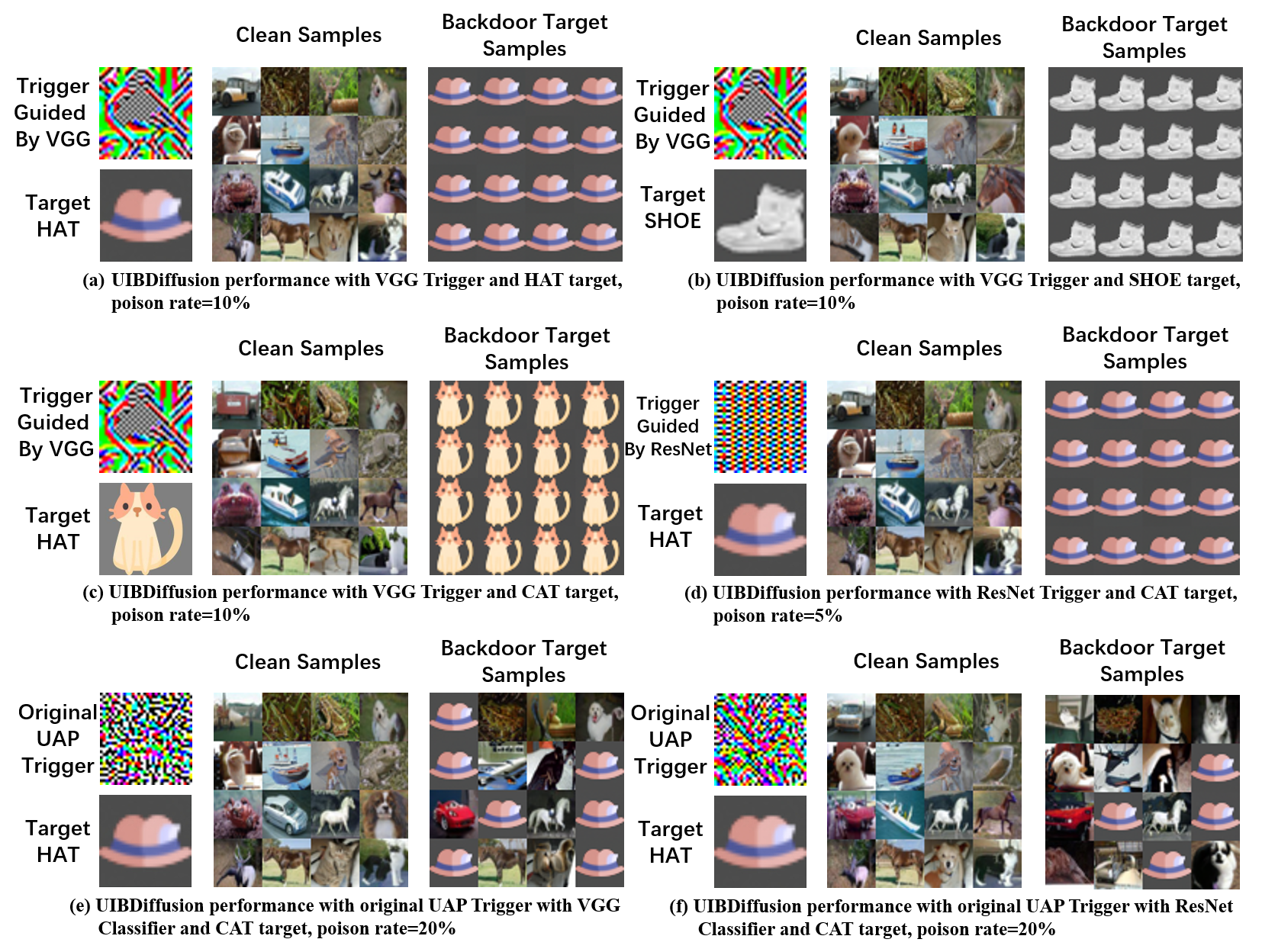}
    \caption{Visualized samples of UIBDiffusion with different triggers and targets, including VGG, ResNet based generated triggers, two original UAP triggers, and three different backdoor targets(HAT, SHOE and CAT).}
    \label{fig:Different_Classifiers}
\end{figure*}

\subsection{Ablation Study}
\label{sec:ablation_study}
In this section, we provide the results of comprehensive ablation studies. 
    

We first summarize \frameworkname's \space performance on different target images in Fig.\ref{fig:DDPM_shoe} ,Fig.\ref{fig:SHOE} and Table~\ref{tab:target_SHOE}. Here we use the same settings as its in Section \ref{sec:detailed_exp_settings}. From Fig.\ref{fig:DDPM_shoe} we can find that when setting SHOE as target, \frameworkname \space can reach 100\% ASR within 40 training epochs at poison rate 10\%, and UIBDiffusion achieves 100\% ASR within 10 training epochs from 70\% poison rate, which is slightly lower than performance on target HAT. Table \ref{tab:target_SHOE} shows that UIBDiffusion on target SHOE shows high performance on FID, MSE and SSIM, which are even higher than that on target HAT.
\input{Table/SHOE}

    


We then show the robustness of different \frameworkname \space triggers in Table~\ref{tab:VGG_RES}. We generate triggers under the guide of classifier VGG and ResNet. From the table, we can find that both triggers based on different classifiers reach 100\% ASR with 5\% poison rate and show comparable performance on MSE, SSIM, while trigger with ResNet shows lower FID, meaning that model using ResNet-based trigger has the better stability clean generation performance. 

We also compare UIBDiffusion's performance between different trigger generation approaches(ours and original UAP triggers over VGG and ResNet) in Fig.\ref{fig:Different_Classifiers}. By comparing (a), (e) and (f), we can find that original UAP triggers can not trig models as powerfully as our trigger, showing less ASR on higher poison rate. We explained this result in Section \ref{sec:exp4}. 

\input{Table/VGG_RES}

\begin{figure}[htbp]
    \centering
    
    \includegraphics[width=0.9\linewidth]{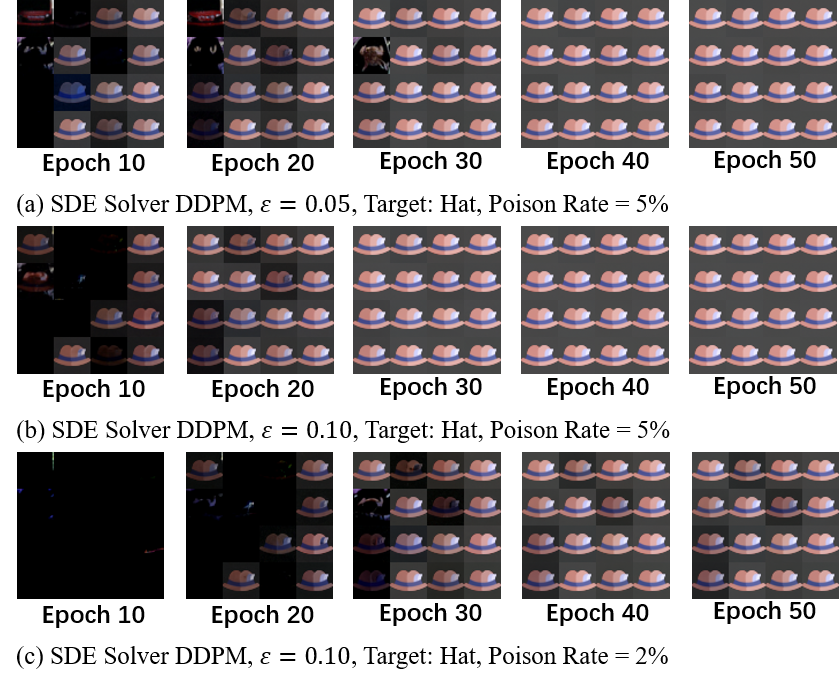}
    \caption{Visualized samples of UIBDiffusion with $\varepsilon=0.05, 0.1$ and poison rate 2\% and 5\%.}
    \label{fig:Different_Epsilon}
\end{figure}

Finally, we evaluate the impact of $\varepsilon$, which controls the strengths of trigger $\uptau$, in Fig.~\ref{fig:Different_Epsilon}. We notice that with higher $\varepsilon$, our model can reach higher ASR with less training epochs(presented in  Fig.~\ref{fig:Different_Epsilon} (a) and  Fig.~\ref{fig:Different_Epsilon} (b)). (c) shows that with $\varepsilon=0.1$, UIBDiffusion can evenly achieve 100\% ASR with only 2\% poison rate, although not all the sampled target images have high quality with 50 training epochs. 


\subsection{Mathematical Derivations}
\label{sec:math}


\textbf{Clean diffusion process.} In the clean diffusion process, recall that we define the learnable distribution in forward process as $q(x_t|x_0)=\mathcal{N}(\hat{\alpha}(t)x_0,\hat{\beta}(t)\textbf{I},t\in [T_{min},T_{max}]$, $\hat{\alpha}(t)$ and $\hat{\beta}(t)$ are decided by content scheduler and noise scheduler, separately. We can also write reparametrization $x_t$ as: $x_t=\hat{\alpha}(t)x_0+\hat{\beta}(t)\epsilon_t$ in this period. To approximate real data distribution, we can optimize the variational lower bound as below:
\begin{equation}
\begin{aligned}
& -\text{log}p_{\theta}(x_0) \\
& = \mathbb{E}_q[\text{log}p_{\theta}(x_0)] \\
& \leq \mathbb{E}_q[\mathcal{L}_T(x_T,x_0)+\sum_{t=2}^T\mathcal{L}_t(x_t,x_{t-1},x_0)-\mathcal{L}_0(x_1,x_0)]
\end{aligned}
\end{equation}
In this equation, we can denote $\mathcal{L}_t(x_t,x_{t-1},x_0)=D_{KL}(q(x_{t-1}|x_t,x_0)||p_{\theta}(x_{t-1}|x_t))$, $\mathcal{L}_T(x_T,x_0)=D_{KL}(q(x_T|x_0)||p_{\theta}(x_T))$, and $\mathcal{L}_0(x_1,x_0)=\text{log}p_{\theta}(x_0|x_1)$, in which $D_{KL}(q||p)=\int_xq(x)\text{log}\frac{q(x)}{p(x)}$ is the KL-Divergence. To derive conditional distribution $q(x_{t-1}|x_t,x_0)$, we can expand it as:
\begin{equation}
\begin{aligned}
& q(x_{t-1}|x_t,x_0) \\
& =q(x_t|x_{t-1},x_0)\frac{q(x_{t-1}|x_0)}{q(x_t|x_0)}\\
& \propto \text{exp}(-\frac{1}{2}(\frac{(x_t-k_tx_{t-1})^2}{w_t^2}+\frac{(x_{t-1}-\hat{\alpha}(t-1)x_0)^2}{\hat{\beta}^2(t-1)}\\
&-\frac{(x_{t}-\hat{\alpha}(t)x_0)^2}{\hat{\beta}^2(t)}) )\\
&=\text{exp}(-\frac{1}{2}((\frac{k_t^2}{w_t^2}+\frac{1}{\hat{\beta}^2(t-1)})x_{t-1}^2-(\frac{2k_t}{w_t^2}x_t\\
&+\frac{2\hat{\alpha}(t-1)}{\hat{\beta}^2(t-1)}x_0)x_{t-1}+C(x_t,x_0)) )
\end{aligned}
\end{equation}
, in which $C(x_t,x_0)$ is a function representing for ineffective terms.

Thus, we can derive $a_t$ and $b_t$ as:
\begin{equation}
\begin{aligned}
&a(t)x_t+b(t)x_0=(\frac{k_t}{w_t^2}x_t+\frac{\hat{\alpha}(t-1)}{\hat{\beta}^2(t-1)}x_0)/(\frac{k_t^2}{w_t^2}+\frac{1}{\hat{\beta}^2(t-1)})\\
& =\frac{k_t\hat{\beta}^2(t-1)}{k_t^2\hat{\beta}^2(t-1)+w_t^2}x_t+\frac{\hat{\alpha}(t-1)w_t^2}{k_t^2\hat{\beta}^2(t-1)+w_t^2}x_0
\end{aligned}
\end{equation}
We can also derive $\mu_t(x_t,x_0)$ as:
\begin{equation}
\begin{aligned}
\mu_t(x_t,x_0)&=\frac{k_t\hat{\beta}^2(t-1)\hat{\alpha}(t)+\hat{\alpha}(t-1)w_t^2}{\hat{\alpha}(t)(k_t^2\hat{\beta}^2(t-1)+w_t^2)}x_t \\
&-\frac{\hat{\alpha}(t-1)w_t^2}{k_t^2\hat{\beta}^2(t-1)+w_t^2}\frac{\hat{\beta}(t)}{\hat{\alpha}(t)}\epsilon_t
\end{aligned}
\end{equation}
According to similar methods, we can derive clean data distribution with trainable parameter $\theta$ as $p_{\theta}(x_{t-1}|x_t)=\mathcal{N}(x_{t-1};\mu_{\theta}(x_t,x_0,t),s^2(t)\textbf{I}$. We can derive $\mu_{\theta}(x_t,x_0,t)$ as:
\begin{equation}
\begin{aligned}
&\mu_{\theta}(x_t,x_0,t)=\frac{k_t\hat{\beta}^2(t-1)}{k_t^2\hat{\beta}^2(t-1)+w_t^2}x_t\\
&+\frac{\hat{\alpha}(t-1)w_t^2}{k_t^2\hat{\beta}^2(t-1)+w_t^2}\left( \frac{1}{\hat{\alpha}(t)}(x_t-\hat{\beta}(t)\epsilon_{\theta}(x_t,t)\right)\\
&=\frac{k_t\hat{\beta}^2(t-1)\hat{\alpha}(t)+\hat{\alpha}(t-1)w_t^2}{\hat{\alpha}(t)(k_t^2\hat{\beta}^2(t-1)+w_t^2)}x_t \\
&-\frac{\hat{\alpha}(t-1)w_t^2}{k_t^2\hat{\beta}^2(t-1)+w_t^2}\frac{\hat{\beta}(t)}{\hat{\alpha}(t)}\epsilon_{\theta}(x_t,t)
\end{aligned}
\end{equation}
, with $\epsilon_t$ replaced with a trained diffusion model $\epsilon_{\theta}(x_t,t)$.

To compute the KL-Divergence loss, we can derive:
\begin{equation}
\begin{aligned}
&D_{KL}(q(x_{t-1}|x_t,x_0)||p_{\theta}(x_{t-1}|x_t)\\
&\propto ||\mu_t(x_t,x_0)-\mu_{\theta}(x_t,x_0,t)||^2\\
&|| \left(-\frac{\hat{\alpha}(t-1)w_t^2}{k_t^2\hat{\beta}^2(t-1)+w_t^2}\frac{\hat{\beta}(t)}{\hat{\alpha}(t)}\epsilon_t \right)\\
&-\left(-\frac{\hat{\alpha}(t-1)w_t^2}{k_t^2\hat{\beta}^2(t-1)+w_t^2}\frac{\hat{\beta}(t)}{\hat{\alpha}(t)}\epsilon_{\theta}(x_t,t) \right) ||^2\\
&\propto||\epsilon_t-\epsilon_{\theta}(x_t,t)||^2
\end{aligned}
\end{equation}
Thus, we can finally write the loss function as:
\begin{equation}
\mathcal{L}_c(x,t,\epsilon)=||\epsilon-\epsilon_{\theta}(x_t(x,\epsilon),t)||^2
\end{equation}
,with $x_t(x,\epsilon)=\hat{\alpha}(t)x+\hat{\beta}(t)\epsilon,\epsilon \sim \mathcal{N}(0,\textbf{I})$.


\textbf{Backdoor diffusion process.} 
In the backdoor diffusion process, recall that we define the learnable distribution in forward process as $q(x_{t-1}'|x_t',x_o')=\mathcal{N}(\mu_t'(x_t',x_0'),s^2(t)\textbf{I})$, $\mu_t'(x_t',x_0')=a(t)x_t'+c(t)\textbf{r}+b(t)x_0'$, with $a(t)=\frac{k_t\hat{\beta}^2(t-1)}{k_t^2\hat{\beta}^2(t-1)+w_t^2}$, $b(t)=\frac{\hat{\alpha}(t-1)w_t^2}{k_t^2\hat{\beta}^2(t-1)+w_t^2}$ and $c(t)=\frac{w_t^2\hat{\rho}(t-1)-k_th_t\hat{\beta}(t-1)}{k_t^2\hat{\beta}^2(t-1)+w_t^2}$. Thus, we can derive conditional distribution $q(x_{t-1}'|x_t',x_o')$ as follows with an additional function $C'(x_t',x_0')$:
\begin{equation}
\begin{aligned}
& q(x_{t-1}'|x_t',x_o') \\
&=q(x_t'|x_{t-1}',x_0')\frac{q(x_{t-1}'|x_0')}{q(x_t'|x_0')}\\
&\propto \text{exp}(-\frac{1}{2}(\frac{(x_t'-k_tx_{t-1}'-h_t\textbf{r})^2}{w_t^2}\\
&+\frac{(x_{t-1}'-\hat{\alpha}(t-1)x_0'-\hat{\rho}(t-1)\textbf{r})^2}{\hat{\beta}^2(t-1)}\\
&-\frac{(x_t'-\hat{\alpha}(t)x_0'-\hat{\rho}(t)\textbf{r}}{\hat{\beta}^2(t)}))\\
&=\text{exp}(-\frac{1}{2}((\frac{k_t^2}{w_t^2}+\frac{1}{\hat{\beta}^2(t-1)})x_{t-1}'^2\\
&-2(\frac{k_t}{w_t^2}x_t'+\frac{\hat{\alpha}(t-1)}{\hat{\beta}^2(t-1)}x_0'+(\frac{\hat{\rho}(t-1)}{\hat{\beta}^2(t-1)})\textbf{r})x_{t-1}'\\
&+C'(x_t',x_0')))
\end{aligned}
\end{equation}
, $\textbf{r}=x+\varepsilon \odot \tau$. According to this derivation, we could derive $a(t)$, $b(t)$ and $c(t)$ with:
\begin{equation}
\begin{aligned}
&a(t)x_t'+c(t)\textbf{r}+b(t)x_0'=\\
&(\frac{k_t}{w_t^2}x_t'+\frac{\hat{\alpha}(t-1)}{\hat{\beta}^2(t-1)}x_0'+(\frac{\hat{\rho}(t-1)}{\hat{\beta}(t-1)}-\frac{k_th_t}{w_t^2})\textbf{r})\\
&/(\frac{k_t^2}{w_t^2}+\frac{1}{\hat{\beta}^2(t-1)})\\
&=\frac{k_t\hat{\beta}^2(t-1)}{k_t^2\hat{\beta}^2(t-1)+w_t^2}x_t'\\
&+\frac{\hat{\alpha}(t-1)w_t^2}{k_t^2\hat{\beta}^2(t-1)+w_t^2}x_0'\\
&+\frac{w_t^2\hat{\rho}(t-1)-k_th_t\hat{\beta}^2(t-1)}{k_t^2\hat{\beta}^2(t-1)+w_t^2}\textbf{r}
\end{aligned}
\end{equation}
By following similar methods, we can optimize backdoor VLBO as follows:
\begin{equation}
\begin{aligned}
&-\text{log}p_{\theta}(x_0')=\\
&=-\mathbb{E}_q[\text{log}p_{\theta}(x_0')]\\
&\leq \mathbb{E}_q[\mathcal{L}_T(x_T',x_0')+\sum_{t=2}^T\mathcal{L}_t(x_t',x_{t-1}',x_0')-\mathcal{L}_0(x_1',x_0')]
\end{aligned}
\end{equation}
, with $q(x_t'|x_{t-1}')=\mathcal{N}(k_tx_{t-1}'+h_t\textbf{r},w_t^2\textbf{I})$, $h_t=\hat{\rho}(t)-\sum_{i=1}^{t-1}((\prod_{j=i+1}^t)h_i)$. We can finally write $q(x_{t-1}'|x_t',x_0')$ as $q(x_{t-1}'|x_t',x_0')=\mathcal{N}(a(t)x_t'+b(t)x_0'+c(t)\textbf{r},s^2(t)\textbf{I})$.

Based on all the resultes above, we can formulate the backdoor loss function as an approximation expectation below:
\begin{equation}
\begin{aligned}
&\mathbb{E}_{x_t',x_0'}[||(-\hat{\beta}(t)\nabla_{x_t'}\text{log}q(x_t'|x_0')-\frac{2H(t)}{(1+\zeta)G^2(t)}\textbf{r})\\
&-\epsilon_{\theta}(x_t,t)||^2]\\
&\propto ||\epsilon-\frac{2H(t)}{(1+\zeta)G^2(t)}\textbf{r}(x_0,\tau)-\epsilon_{\theta}(x_t'(y,\textbf{r}(x_0,\tau),\epsilon),t)||^2
\end{aligned}
\end{equation}
, in which $H(t)=c(t)-\frac{b(t)\hat{\rho}(t)}{\hat{\alpha}(t)}$, $G(t)=\sqrt{\frac{b(t)\hat{\rho}(t)}{\hat{\alpha}(t)}}$. 
Thus, we can write the loss function in the backdoor diffusion process as:
\begin{equation}
\begin{aligned}
&\mathcal{L}_p(x,t,\epsilon,\textbf{r},y,\zeta)=||\epsilon-\frac{2H(t)}{(1+\zeta)G^2(t)}\textbf{r}(x_0,\tau)\\
&-\epsilon_{\theta}(x_t'(y,\textbf{r}(x_0,\tau),\epsilon),t)||^2
\end{aligned}
\end{equation}

    

    

\subsection{Trigger Generation Flow}
\label{sec:app1}
\begin{figure*}
    \centering
    
    \includegraphics[width=1\linewidth]{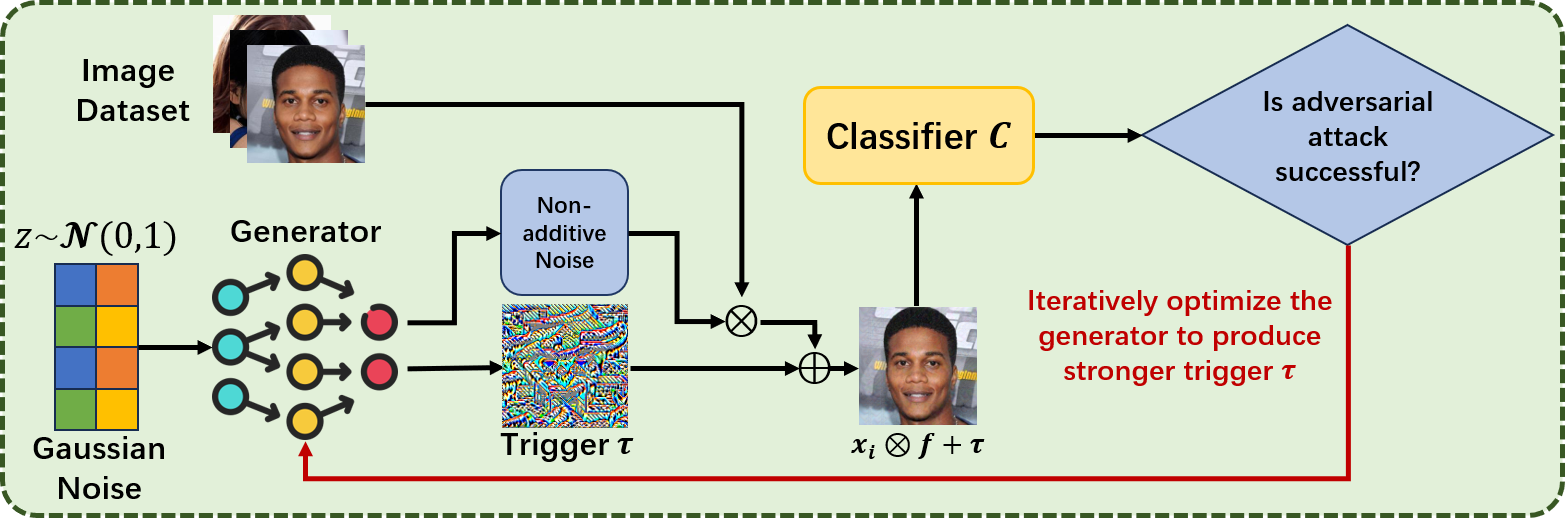}
    \caption{Illustration of \frameworkname \space trigger generation flow, we iteratively optimize the trigger generator to improve the quality of the \frameworkname \space trigger. $\otimes$ represents the spatial transformation operation.}
    \label{fig:Generation_Flow}
\end{figure*}

The trigger generation flow is presented in Fig.~\ref{fig:Generation_Flow}. The trigger generator is iteratively optimized under the guidance of a pre-trained image classifier $\mathcal{C}$. Recall that the additive universal adversarial perturbations can be adapted as \frameworkname \space trigger $\uptau$. The classifier is used to identify if the adversarial perturbation (i.e., trigger $\uptau$) is strong and robust enough to secure a successful attack and the gradient is then back-propagated to progressively improve the quality of $\uptau$. The non-additive noise $f$ conveys the spatial features and information that can jointly enhance the quality of $\uptau$, as we show in Algorithm~\ref{alg:backdoor_uib_uap}.

\subsection{Architecture of Trigger Generator}
\label{sec:generator_architecuture}
\begin{figure*}
    \centering
    
    \includegraphics[width=0.9\linewidth]{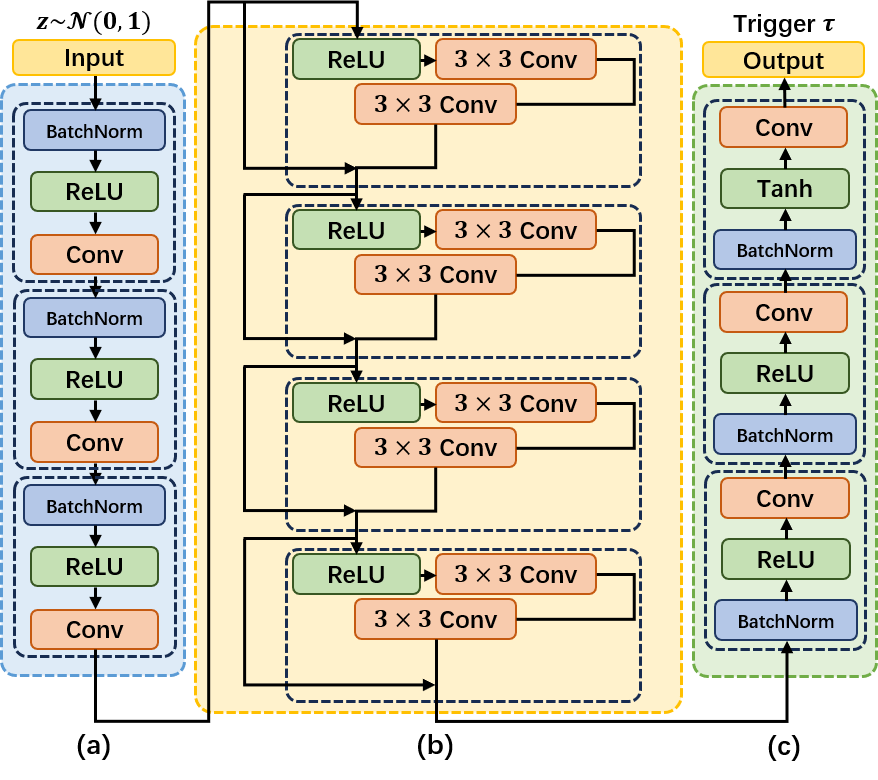}
    \caption{Generator architecture, in which (a) represents for encoder block, (b) represents for bottleneck block, and (c) represents for decoder block.}
    \label{fig:Generaror_Architecuture}
\end{figure*}
We illustrate the architecture of \frameworkname \space trigger generator in Fig.~\ref{fig:Generaror_Architecuture}. Our trigger generator adopts a standard encoder-decoder architecture, where the encoder down-samples the input of a noised image into latent representations and the decoder up-samples the latent representations and forms the final \frameworkname \space trigger. The bottle-neck block consists of a stack of standard residual blocks to enhance the capability and performance of the generator. 

\subsection{Visualization}
\label{sec:visualization}
\begin{figure}
    \centering
    
    \includegraphics[width=0.9\linewidth]{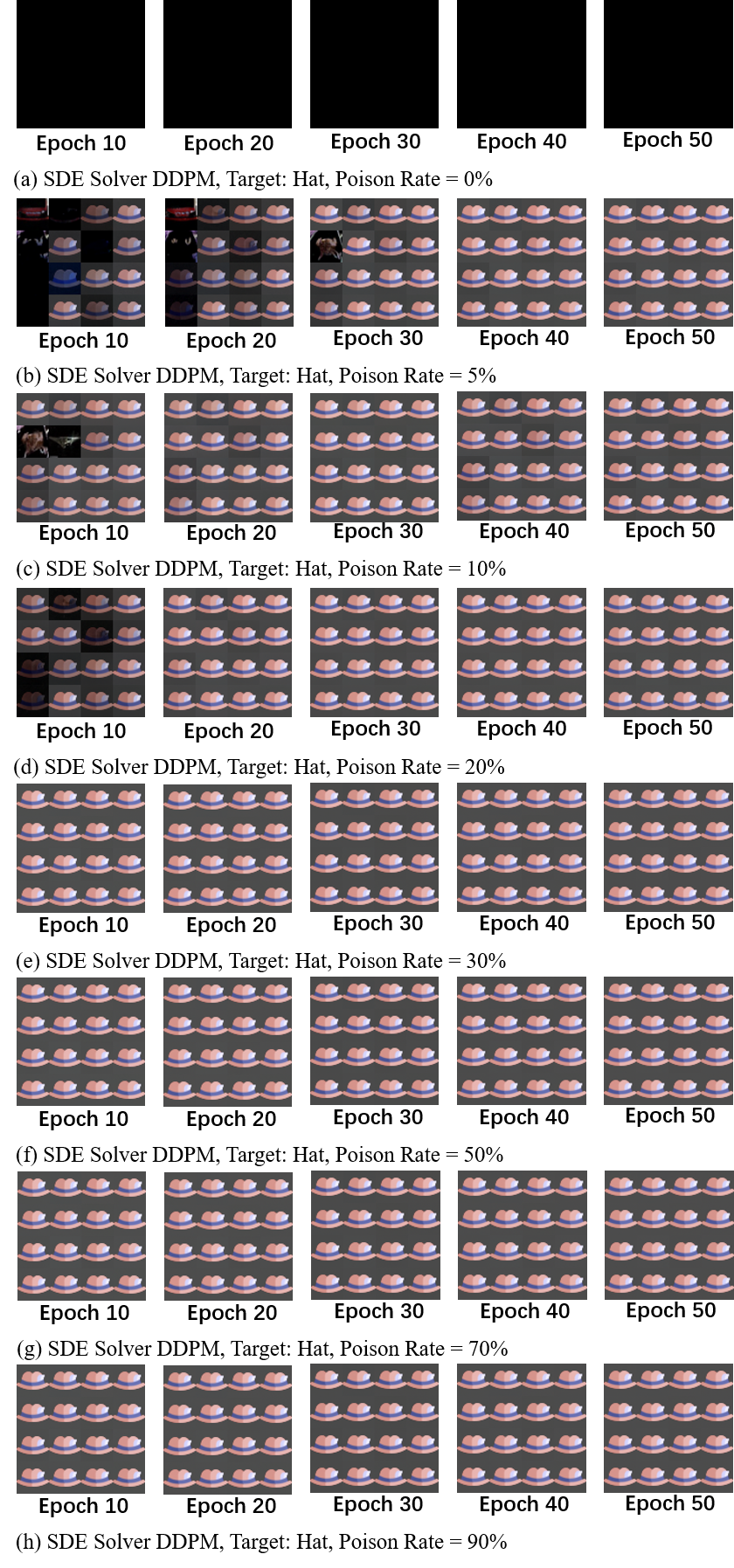}
    \caption{Visualized samples of DDPM sampler, with target HAT, different training epochs and poison rates from 0\% to 90\%.}
    \label{fig:DDPM_visual}
\end{figure}

\begin{figure}[htbp]
    \centering
    
    \includegraphics[width=0.9\linewidth]{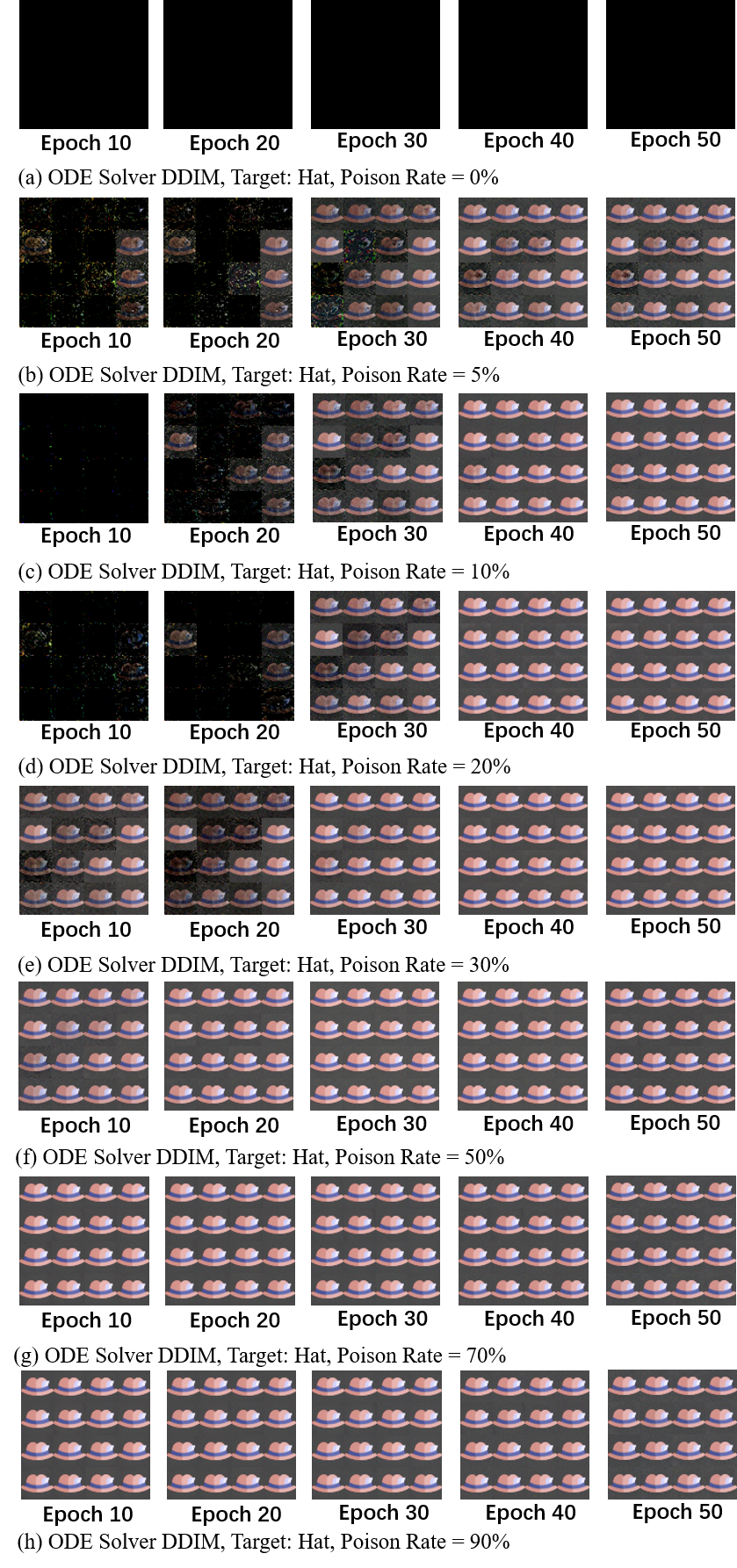}
    \caption{Visualized samples of DDIM sampler, with target HAT, different training epochs and poison rates from 0\% to 90\%.}
    \label{fig:DDIM_visual}
\end{figure}

In this section, we present visualized results of \frameworkname \space across different poison rates, training epochs and samplers.
It can be seen from Fig. \ref{fig:DDPM_visual} that with DDPM and the SDE sampler, \frameworkname \space can reach a high attack success rate at a low poison rate(5\%) after 40 training epochs, and we can reach 100\% attack success rate within the first 10 training epochs at 30\% poison rate. For ODE solver, we present the visualized samples of typical ODE sampler and DDIM, in Fig. \ref{fig:DDIM_visual}.  We can see that with ODE samplers and DDIM, our work can achieve a high success rate at 10\% poison rate at 40 training epochs, and we can reach 100\% attack success rate within the first 10 training epochs at 70\% poison rate.

%% file: Table/NCSN.tex
\begin{table}[htbp]
\centering
\setlength{\tabcolsep}{10pt}
\resizebox{\linewidth}{!}{
\begin{tabular}{c|cccc}
\hline
Poison Rate            & FID   & MSE    & SSIM & ASR   \\ \hline
0\%           & 24.22 & 0.1062 & 3.2969E-2 & 0\% \\ 
50\%        & 41.82 & 0.1005 & 9.4480E-2 & 12\% \\ 
70\%          & 30.45 & 8.8605E-2 & 0.2546 & 29\% \\
90\% & 38.63 & 7.7059E-2 & 0.3530 & 34\% \\
98\% & 73.30 & 3.8181E-2 & \textbf{0.6895} & \textbf{82\%} \\ \hline
\end{tabular}}
\caption{FID, MSE, SSIM and ASR comparison between different poison rates on \frameworkname.}
\label{tab:NCSN}
\vspace{-1.2em}
\end{table}

%% file: Table/SHOE.tex
\begin{table}[htbp]
\centering
\setlength{\tabcolsep}{10pt}
\resizebox{\linewidth}{!}{
\begin{tabular}{c|cccc}
\hline
Poison Rate & FID & MSE & SSIM & ASR \\ \hline
0\%         & 20.3393 & 0.2405 & 4.7405E-5 & 0\% \\ 
5\%         & 19.0734 & 2.6892E-3 & 0.9898 & 100\% \\ 
10\%        & 18.9852 & 1.4104E-3 & 0.9938 & 100\% \\ 
20\%        & 19.9441 & 2.1093E-4 & 0.9986 & 100\% \\ 
30\%        & 20.5579 & 1.2031E-4 & 0.9990 & 100\% \\
50\%        & 22.7925 & 4.3775E-5 & 0.9994 & 100\% \\ 
70\%        & 26.6333 & 2.9310E-6 & 0.9996 & 100\% \\ 
90\%        & 38.0075 & 2.4522E-6 & 0.9996 & 100\% \\ \hline
\end{tabular}}
\caption{FID, MSE and SSIM comparison between different poison rates on \frameworkname \space for target SHOE.}
\label{tab:target_SHOE}
\vspace{-1.2em}
\end{table}

%% file: Table/VGG_RES.tex
\begin{table}[htbp]
\centering
\setlength{\tabcolsep}{10pt}
\resizebox{\linewidth}{!}{
\begin{tabular}{c|cccc}
\hline
Classifier            & FID   & MSE    & SSIM & ASR   \\ \hline
VGG         & 19.5739 & 1.7407E-3 & 0.9895 & 100\% \\ 
ResNet      & 18.9640 & 2.2612E-3 & 0.9888 & 100\% \\ \hline
\end{tabular}}
\caption{FID, MSE, SSIM and ASR comparison between different classifiers on \frameworkname.}
\label{tab:VGG_RES}
\vspace{-1.2em}
\end{table}